\documentclass[reprint,aps,prl,nofootinbib,
superscriptaddress
]{revtex4-1}
\usepackage{graphicx}
\usepackage{dcolumn}
\usepackage[svgnames]{xcolor}
\usepackage{booktabs}
\usepackage{multirow}
\usepackage{makecell}
\usepackage{hyperref}
\newcommand{\LG}[2]{${\rm LG}_{#1}^{#2}$}
\newcommand{\CM}[1]{${\rm LG}_0^{|{#1}|}$}

\begin{document}

\title{Orbital angular momentum beam generation using a free-electron laser oscillator}

\author{Peifan Liu}
\author{Jun Yan}
\affiliation{FEL Laboratory, TUNL and Department of Physics, Duke University, Durham, NC 27708, USA}
\author{Andrei Afanasev}
\affiliation{Department of Physics, The George Washington University, Washington, D.C. 20052, USA}
\author{Stephen V. Benson}
\affiliation{Thomas Jefferson National Accelerator Facility, Newport News, VA 23606, USA}
\author{Hao~Hao}
\author{Stepan F. Mikhailov}
\author{Victor G. Popov}
\author{Ying K. Wu}
\email{wu@phy.duke.edu}
\affiliation{FEL Laboratory, TUNL and Department of Physics, Duke University, Durham, NC 27708, USA}


\begin{abstract}
With wavelength tunability, free-electron lasers (FELs) are well-suited for generating orbital angular momentum (OAM) beams in a wide photon energy range.
We report the first experimental demonstration of OAM beam generation using an oscillator FEL.
Lasing around $458$~nm, we have produced the four lowest orders of coherently mixed OAM beams 
with good beam quality, excellent stability, and substantial intracavity power. 
We have also developed a pulsed mode operation of the OAM beam with a highly reproducible temporal structure for a range of modulation frequencies from $1$ to $30$~Hz.
This development can be extended to short wavelengths, for example to x-rays using a future x-ray FEL oscillator.
The operation of such an OAM FEL also paves the way for the generation of OAM gamma-ray beams via Compton scattering. 
\end{abstract}

\maketitle

The profound and unexpected properties of optical beams with orbital angular momentum (OAM) were first recognized in the early 1990s~\cite{allen1992orbital,beijersbergen1993astigmatic,allen1999iv}. 
In the last two decades, OAM photon beams in the near-infrared and visible regimes have been generated using multiple methods based on conventional laser technology~\cite{he1995optical,beijersbergen1994helical,ito2010generation,naidoo2016controlled}.
Recent research has shown that the OAM photon beam is an excellent tool for non-contact optical manipulation of matter.
A wide range of applications have been discovered ~\cite{franke2008advances,yao2011orbital,rubinsztein2016roadmap}, from biological cell handling in optical tweezers~\cite{simpson1996optical,grier2003revolution,dasgupta2011optical}, to laser cooling, atom trapping and control of Bose-Einstein condensates~\cite{kuga1997novel,tabosa1999optical,andersen2006quantized,wright2008optical}, and to quantum information and quantum communication~\cite{molina2001management,gibson2004free,nagali2009quantum,vallone2014free}. 
The OAM photon beam has also been 
shown to excite forbidden transitions in an atom~\cite{schmiegelow2016transfer,afanasev2018experimental,picon2010photoionization}, overcoming the transition selection rules associated with plane-wave photons. 
Many new research opportunities are expected 
with OAM beams at shorter wavelengths, from extreme ultraviolet (EUV) to gamma-ray. EUV and x-ray OAM beams can be used to improve the contrast of microscopy~\cite{sakdinawat2007soft}, 
enable new forms of spectroscopy~\cite{van2007prediction}, 
or alter material magnetic properties \cite{watzel2016optical,fujita2017ultrafast}.
OAM gamma rays may open new possibilities in photo-nuclear physics research.
Theoretically, they have been shown to modify photo-nuclear reaction rates~\cite{afanasev2018radiative}, reveal novel spin effects~\cite{afanasev2017circular,ivanov2020doing}, 
provide the means to separate resonances with different spins and parities~\cite{ivanov2020kinematic}, and enable new types of multipole analysis in photo-reactions~\cite{afanasev2018atomic}. 
OAM gamma-rays may have been generated in extreme astrophysics environments~\cite{taira2018generation,katoh2017angular,maruyama2019compton,maruyama2019comptonHG}.
The quantum vortex nature of an OAM photon's wave function
leads to ``superkick'' effects that may be observed both in the laboratory and astrophysics environment~\cite{afanasev2020recoil}, addressing, among other things, the issue of the universe's transparency to very high energy gamma rays.

The generation of OAM photon beams in the short wavelength region can be realized using accelerator-based light sources.
For example, a relativistic electron beam traversing a helical undulator magnet produces 
higher-order harmonic radiation off-axis.
This radiation was first recognized to exhibit an intrinsic spiral wavefront in a theoretical analysis~\cite{SSasaki2008,afanasev2011generation},
a finding confirmed later experimentally~\cite{JBahrdt2013}.
Recent advances 
have led to the generation of an OAM beam in several single-pass free-electron lasers (FELs) seeded using a laser in the fundamental Gaussian mode~\cite{EHemsing2013,EHemsing2014,PRibivc2017}.
The laser interaction with the electron beam in an upstream undulator (a modulator) modulates the electron beam energy distribution and this energy modulation is turned into a charge density modulation. 
With a helical modulator, an electron beam with helical microbunching can be used to produce an OAM beam in a downstream undulator (a radiator)~\cite{EHemsing2013}.
With a planar modulator, the electron beam with longitudinal microbunching
can be used to produce an OAM beam via harmonic radiation
in the downstream helical radiator, either via a straightforward second-harmonic generation~\cite{EHemsing2014}, or
using a high-gain harmonic generation scheme~\cite{LYu2000} to up-shift the radiation wavelength to the EUV~\cite{PRibivc2017}.

\begin{figure}[t!]
\centering
\includegraphics[width=\columnwidth]{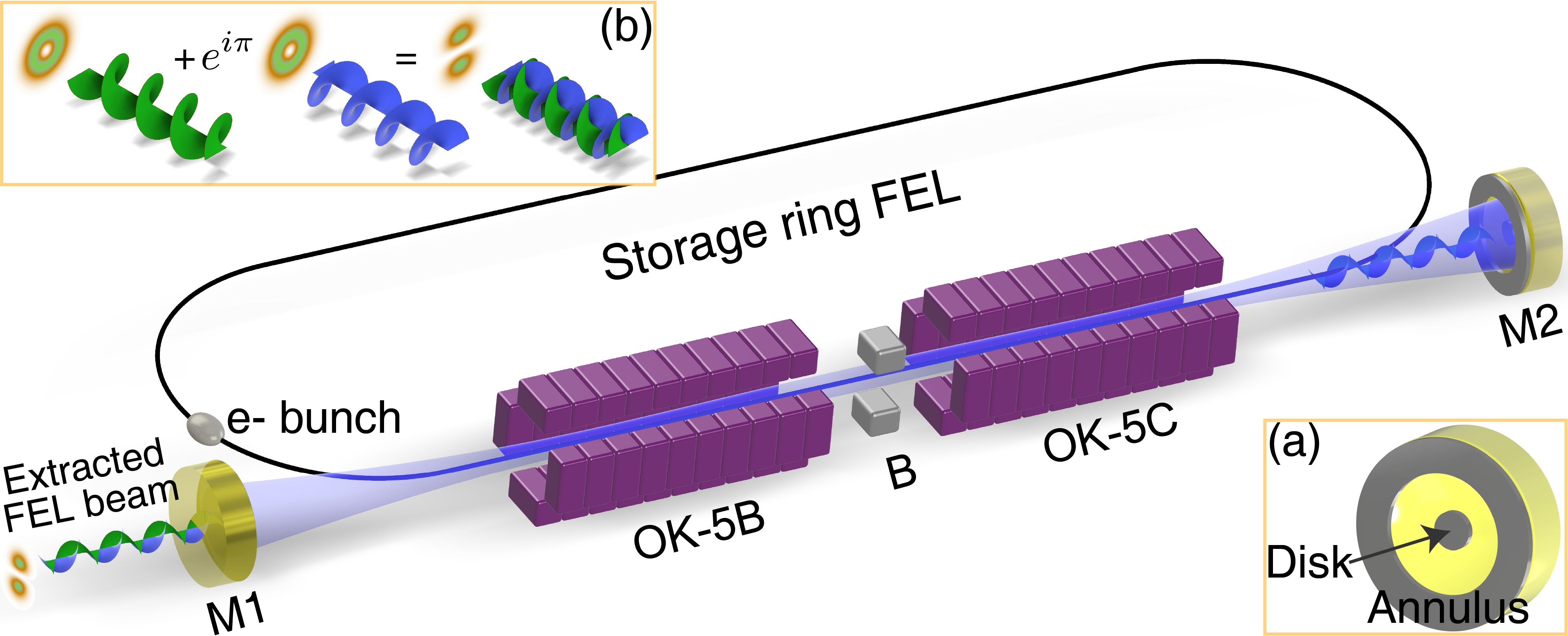}
\caption{
Typical schematic layout of the storage ring FEL oscillator for the OAM beam generation. The FEL is comprised of an optical klystron with two helical undulators OK-5B and OK-5C and a buncher magnet B in between, and a near-concentric optical resonator with mirrors M1 and M2.
Downstream mirror M2 is covered with a special spatial mask with two non-reflective parts [see the bottom-right inset (a)]: an aluminum disk in the center and an aluminum annulus covering the exterior region of the mirror.
The top-left inset (b) illustrates the coherent mixing of the \LG{0}{1} and \LG{0}{-1} modes, which have the same intensity profile but different wavefront structures.
}
\label{Fig:Layout}
\end{figure}

In this Letter, we report the first experimental realization of OAM laser beams using an oscillator FEL.
In this work, a self-seeded OAM beam is amplified in multiple passes inside a laser resonator to reach saturation, unlike the previous OAM beam generation using seeded single-pass FELs~\cite{EHemsing2013,EHemsing2014,PRibivc2017}.
An oscillator FEL is well suited for producing high intracavity power and high-repetition-rate coherent radiation in a wide wavelength range, with demonstrated operation from the far infrared to the vacuum UV~\cite{MBillardon1983,NVinokurov1989,FGlotin1993,STakano1993,GNeil2000,VLitvinenko2001,OShevchenko2016}. 
The schematic layout of the Duke FEL is
shown in Fig.~\ref{Fig:Layout}.
Powered by an electron storage ring, this FEL is comprised of a  near-concentric, $53.73$~m long optical resonator and a magnetic system with multiple undulator magnets~\cite{wu2006high}.
In a typical FEL configuration shown in Fig.~\ref{Fig:Layout}, two middle helical undulators, OK-5B and OK-5C, are energized; these undulators, together with a buncher magnet B sandwiched between them form an optical klystron (OK)~\cite{vinokurov1977report,litvinenko1998first} to enhance the FEL gain.

When operating at its fundamental frequency, the FEL wavelength can be tuned by changing the electron beam energy ($E_e$) and the undulator magnetic field according to the formula: $\lambda_{\rm FEL} = {\lambda_u} \left(1 + {K^2}\right)/(2\gamma^2)$.
Here, $\gamma =E_e/m_ec^2$, $K = eB_0\lambda_u/(2\pi m_ec)$
the undulator strength, with $m_e$ being the electron's rest mass, $\lambda_u$ the undulator period,
$B_0$ the rms magnetic field, and $c$ the speed of light.
After repeated electron-photon beam interactions, the FEL beam builds up inside the cavity between two high-reflectivity mirrors.
The typical FEL transverse mode is the lowest Gaussian mode, 
realized with good alignment of the electron beam orbit and optical axis.
To generate OAM beams in the Laguerre-Gaussian \LG{p}{l} modes (with radial order $p$ and OAM order $l$), 
the cylindrical symmetry of the cavity's transverse boundary must be assured. We accomplish this by using a spatial mask inside the cavity. 

This spatial mask is integrated with the downstream mirror M2 [see inset (a) of Fig.~\ref{Fig:Layout}]. 
The mask consists of an aluminum central disk and an aluminum annulus.
With a carefully chosen inner diameter ($D_{\rm annu}$), the annulus enforces the cylindrical symmetry of the exterior cavity boundary.
The high loss central disk (diameter $D_{\rm disk}$) is used to prevent the formation of the fundamental and other low-order modes.
Multiple disks are fabricated with their dimensions estimated by comparing the loss of a particular mode with a set of possible FEL gain values.
For all experimental results with various orders of OAM beams, a fixed $D_{\rm annu} =26$~mm is used, while $D_{\rm disk}$ is varied with the order of OAM modes to suppress undesirable lower order modes. 
Dimensions of the spatial mask, together with other key beam parameters are provided in Table~\ref{Tab:ExpPara}.
For results reported in this Letter, the typical experimental setup involves a single-bunch electron beam of $533$~MeV and FEL lasing at $\lambda_{\rm FEL}=458$~nm with circular polarization.
Non-typical experimental settings will be noted explicitly. 
The measured electron bunch length varies from about $100$~ps to $210$~ps, depending on the storage ring FEL operating conditions.

\begin{table}[b]
\caption{\label{Tab:ExpPara} Summary of experimental parameters.}
\begin{ruledtabular}
\begin{tabular}{lc}
Electron beam energy $E_e$ (MeV) & $490$--$533$\\
Electron single-bunch current $I_b$ (mA) & $10$--$50$\\
FEL wavelength $\lambda_{\rm FEL}$ (nm) &  $454$--$458$ \\	
Annulus inner diameter $D_{\rm annu}$~(mm)& 26\\
\cmidrule{2-2}
  \multirow{3}{*}{\makecell{Disk diameter $D_{\rm disk}$~(mm)\\(with FEL generated OAM modes)}} 
  & 2.0 (\CM{1}, \CM{2})\\
  & 3.7 (\CM{2}, \CM{3})\\
  & 4.2 (\CM{3}, \CM{4})\\
\end{tabular}
\end{ruledtabular}
\end{table}

For an optical cavity with cylindrical symmetry, two pure OAM 
modes of the same frequency but opposite helicities, \LG{0}{l} and \LG{0}{-l} 
are degenerate---both have the same transverse intensity distribution, but different spiraling orientations of the wavefront.
As a result, a coherent mixing beam \CM{l} is usually produced, with its electric field given by
\begin{equation}
{\rm LG}_0^{|l|}
=\frac{1}{\sqrt{2}}\left(
{{\rm LG}_0^{l}} + e^{i\phi_0} {{\rm LG}_0^{-l}}\right),
\end{equation}
where $\phi_0$ is the phase difference between two base modes.
The intensity profile of this beam, 
featuring 2$l$ azimuthally distributed high intensity regions (petals), is readily used to recognize the order of the underlying OAM modes \cite{pereira1998pinning,chen2001generation,naidoo2012intra,lin2014controlling}.
The phase difference $\phi_0$ determines the azimuthal orientation of the petals.
For an FEL cavity with a spatial mask, 
the FEL beam is always generated in the coherently mixed mode.
In Fig.~\ref{Fig:Layout}, inset (b) illustrates the coherent mixing of \LG{0}{1} and \LG{0}{-1} modes with $\phi_0 = \pi$, i.e. a \CM{1} beam. 
In Table \ref{Tab:ExpPara} the orders of the generated OAM beams are tabulated for the various disk sizes used.

Lasing in the OAM mode is realized by maximizing gain via careful overlap of the electron beam and the optical mode and the use of an optical klystron configuration. 
If necessary a third undulator can be turned on to further enhance the gain. The transverse beam overlap is optimized not only by alignment, but also by increasing the vertical beam size by operating the storage ring near the transverse coupling resonance.
The frequency of the storage ring rf system is tuned carefully to realize
nearly perfect synchronization between the electron and FEL beams (i.e. zero FEL detuning).
With careful tuning of the FEL, OAM beams of several orders have been generated with excellent reproducibility.

\begin{figure}[t]
\centering
\includegraphics[width=\linewidth]{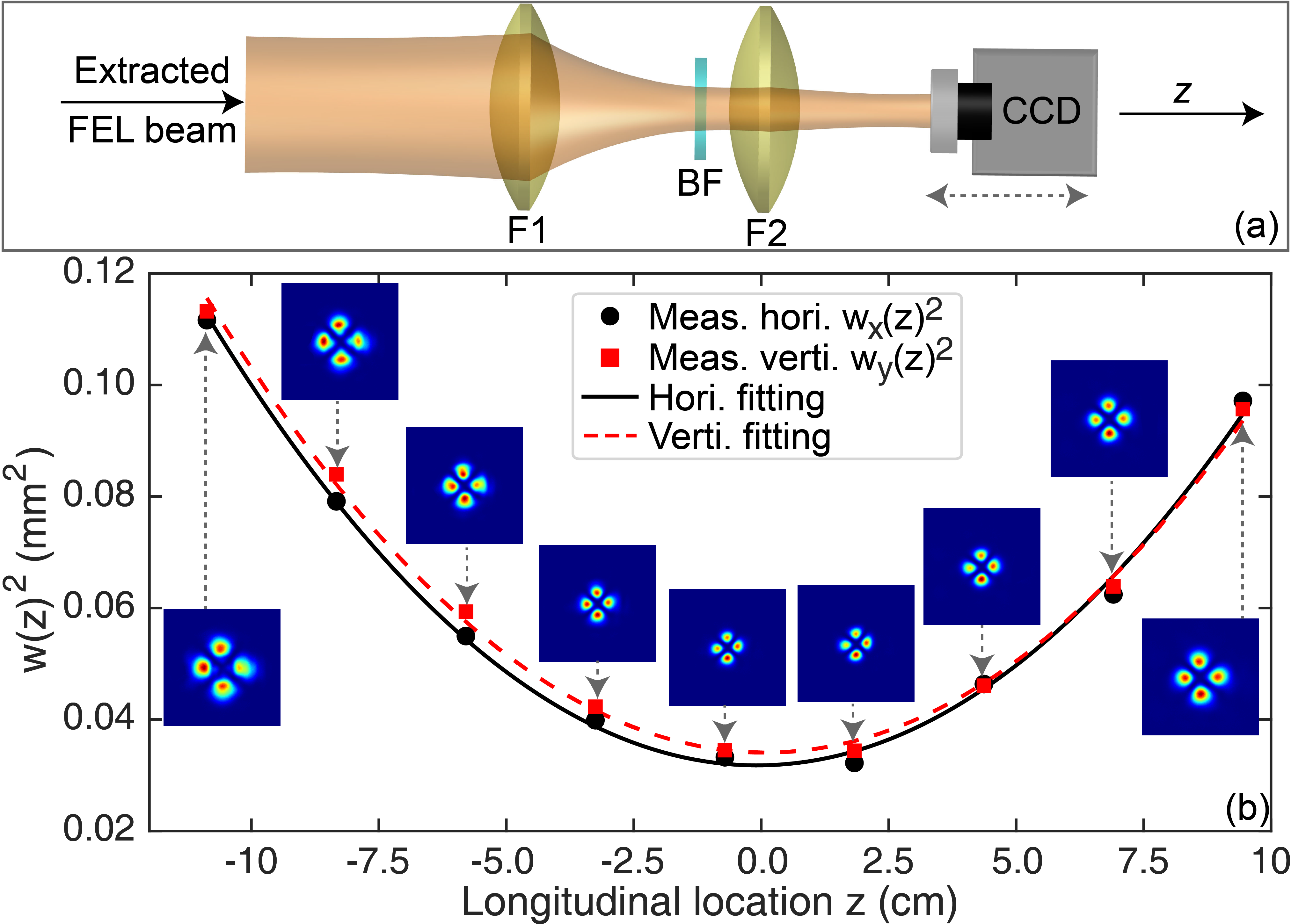}
\caption{(a) Schematic layout for the FEL beam profile measurement system with a movable CCD and a telescopic transport optics comprised of focusing lenses F1 and F2. 
Following F1 is a narrow bandpass filter BF. 
(b) Beam quality factor $M^2$ measurement of the \CM{2} beam. The inserted beam images show the transverse beam profiles at different longitudinal locations ($z$). $M_{x,y}^2$ is determined by fitting the measured square of the beam width, $w_{x,y}^2$, to a quadratic function of $z$,
and $w_{x,y}=2\sigma_{x,y}$ with $\sigma^2_{x,y}$ being the second-order moment of beam's intensity distribution in the horizontal (vertical) direction.
For this measurement, the single-bunch current $I_b\approx 25$~mA.}
\label{Fig:BeamImageM2}
\end{figure}

The extracted FEL beam from upstream mirror M1 is sent to a set of standard FEL diagnostics to measure the beam spectrum, power, etc., and to a new diagnostic to measure the beam profile.
This FEL profile measurement system is comprised of a telescopic beam transport, a narrow bandpass filter, and a charge-coupled device (CCD) camera with 8-bit resolution [see Fig.~\ref{Fig:BeamImageM2}(a)]. 
The optical telescope consists of two focusing lenses F1 and F2 with focal lengths $f_1 =50$~cm and $f_2=10$~cm, respectively.

By moving the CCD camera along the beam direction, the beam profile can be recorded as a function of the longitudinal position ($z$).
To effectively use the camera's dynamic range and avoid saturation, the exposure time of the CCD is adjusted, and when needed, a neutral density filter is inserted before the camera.
During the measurement, the FEL beam power may fluctuate slightly, which is typically corrected by fine-tuning the FEL optical axis or adjusting the FEL detuning. 
A series of beam images along the beam direction is shown in Fig.~\ref{Fig:BeamImageM2}(b)
for the \CM{2} beam.
By using the projected beam distribution in the horizontal ($x$) and vertical ($y$) directions, the beam quality factor $M^2$ can be obtained using a quadratic curve fitting \cite{siegman1990new}. 
For this dataset, 
we find  $M_{x}^2=3.2$ and $M_{y}^2=3.3$.
These values are a factor of $1.07$ and $1.1$ above the theoretical value ($M^2=3$ for the \CM{2} mode), which is one of the best results among multiple \CM{2} mode quality factor measurements. 
The discrepancy in $M^2$ may indicate the presence of a very small amount of other transverse modes in the beam, or some distortion in the optical image due to factors, such as diffraction, noise, etc.

\begin{figure}[t]
\centering
\includegraphics[width=\linewidth]{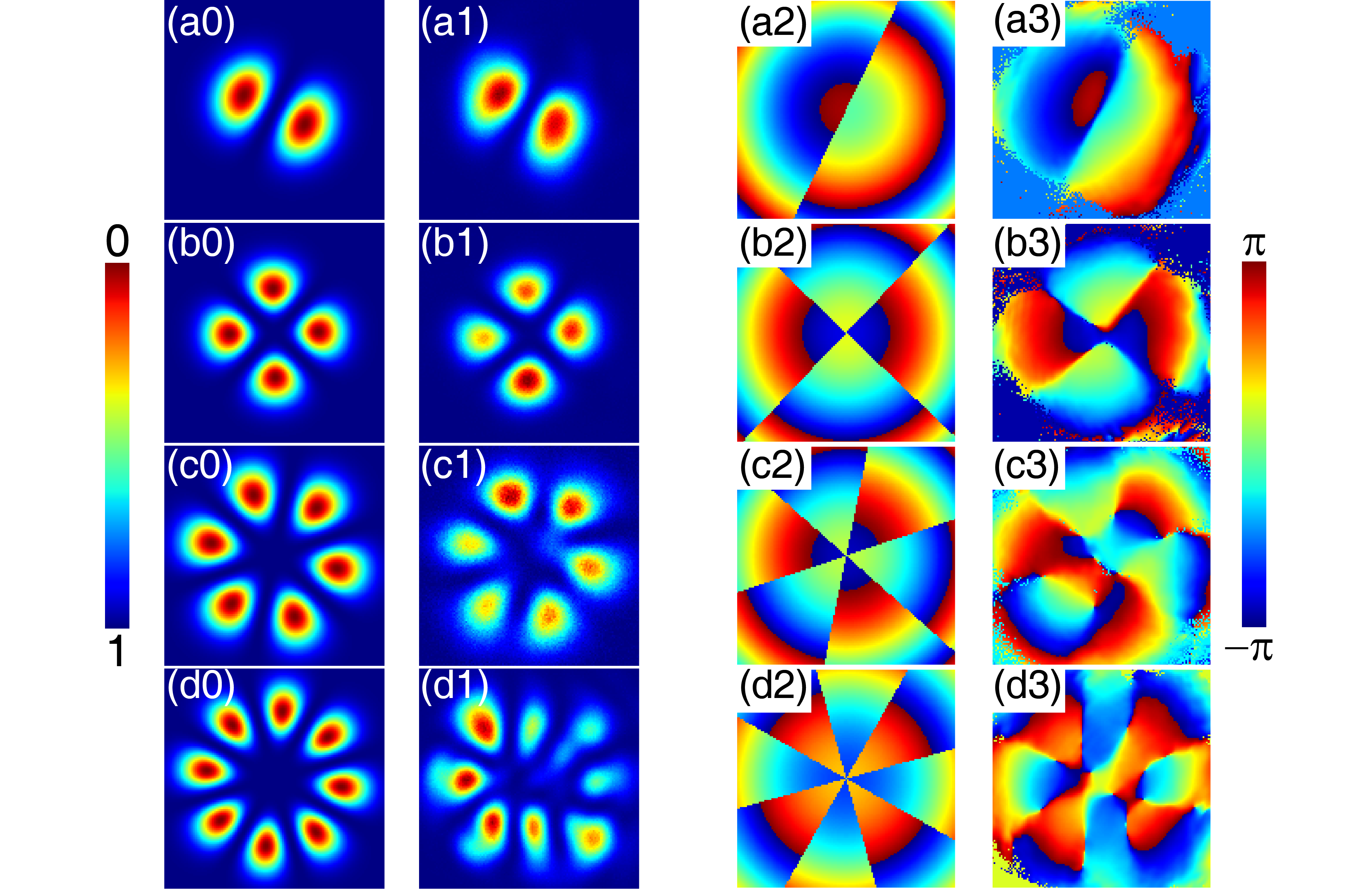}
\caption{
(a0)--(d0) Intensity profiles of theoretical modes \CM{1}, \CM{2}, \CM{3}, \CM{4}, respectively (image sizes: $127\times 127$ pixels). (a1)--(d1) Intensity profiles of the experimentally measured corresponding beams. 
(a2)--(d2) Phase distribution
of corresponding ideal OAM modes. (a3)--(d3) Phase distribution of the measured beams. 
For these measurements, the beam parameters vary, $E_{e}= 490$--$518$~MeV, $I_b=14$--$30$~mA, and $\lambda_{\rm FEL} = 454$--$457$~nm.}
\label{Fig:PhaseRetrieval}
\end{figure}

Using the measured intensity profile at different longitudinal locations, an iterative phase retrieval technique \cite{liu2020phaseret} has been applied to \CM{1}, \CM{2}, \CM{3}, and \CM{4} beams to obtain the relevant wave phase information.
Comparing the 
measured intensity and phase distributions with theoretical ones for these four beams (shown in Fig.~\ref{Fig:PhaseRetrieval}), the measured results of the first three beams agree well with the theoretical ones. 
For the \CM{4} beam, the difference between the theoretical and the measured results is more pronounced, which is possibly due to a small FEL net gain for this large size, higher-order mode due to substantial cavity losses. 

\begin{figure}[t!]
\centering
\includegraphics[width=\linewidth]{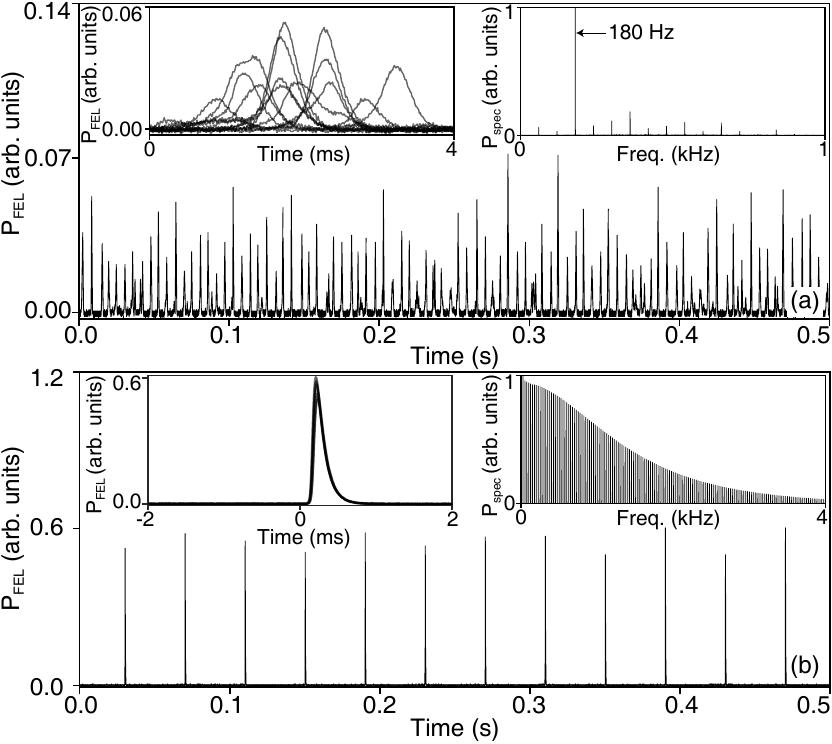}
\caption{
(a) Measured temporal structure of the \CM{2} beam as the FEL is free-running without deliberate modulation; Top-left inset: the first $12$ macropulses in a fixed time window with the time delay associated with $180$~Hz ac line frequency removed; Top-right inset: the related power spectrum of the beam signal, the frequency for the highest peak is $180$~Hz. 
(b) Temporal structure of the \CM{2} beam with a $25$~Hz external modulation; 
Top-left inset: 12 macropulses in a fixed time window with the time delay associated the modulation trigger removed; 
Top-right inset: the related power spectrum. 
For this measurement, the single-bunch current $I_b\approx 30$~mA.
}
\label{Fig:PulsedMode}
\end{figure}

The storage ring FEL beam has a complex temporal structure with both macropulses and micropulses~\cite{billardon1990storage,billardon1992saturation}.
The micropulses are associated with the electron beam revolution ($2.79$~MHz in our case) and the macropulses are the result of 
two competing processes: FEL lasing and radiation damping of the electron beam.
With zero FEL detuning, the FEL lasing in the fundamental Gaussian mode can produce a quasi-CW beam with reasonably stable micropulses without the apparent macropulse structure.
However, the OAM beams behave differently: they always remain in a pulsing mode even with zero detuning.
For example, when free-running without external modulation, the temporal structure of a \CM{2} beam [Fig.~\ref{Fig:PulsedMode}(a)] is dominated by a pulsing frequency of $180$~Hz, the third harmonic of the ac line frequency (see the top-right inset).
This can be considered as a natural modulation frequency due to ac noise in the system.
Besides substantial variations in the pulse shape and height, the arrival time of the well-formed macropulses varies by about $1$~ms (rms) after removing the time delay associated with $180$~Hz  
(see the top-left inset).
For this free-running \CM{2} beam, the maximum extracted power 
is about $14$~mW with a high beam current ($I_b\approx50$~mA).
This corresponds to a conservative estimate of the intracavity laser power of about $10$~W.

To improve the reproducibility of the OAM beam, we used an FEL gain modulator to deliberately modulate the longitudinal coupling between the electron beam and FEL beam.
The FEL can be modulated with a wide range of frequency ($f_{\rm mod}$), and in our measurements, $f_{\rm mod}$ was varied from $1$ to $67$~Hz.
We find that the pulse energy of the OAM beam remains relatively constant for $f_{\rm mod}$ between $1$ and $25$~Hz, and decreases at higher frequencies.
On the other hand, the average FEL power increases linearly with
$f_{\rm mod}$ up to about $25$~Hz and then tapers off at higher frequencies.
A close examination of OAM operation at various modulation frequencies also shows this beam has the best  reproducibility in terms of the energy per macropulse at $25$~Hz.
The temporal structure of the OAM beam at $25$ Hz is shown in Fig.~\ref{Fig:PulsedMode}(b), with very regularly displaced pulses in both the time and frequency domains (the top-right inset).
The macropulses are highly reproducible as illustrated by using the first 12 pulses in the top-left insets of Fig.~\ref{Fig:PulsedMode}(b).
The analysis of the full dataset ($5$ second long) shows that the energy per pulse varies by $\sim4.7$\% (rms), the pulse width varies by $\sim1.7$\% (rms), and pulse arrival time varies by $\sim7$~$\mu$s (rms) determined using the half-height of the pulse front (after removing the modulation trigger delay).
Overall, the pulsed OAM beam operation is optimum when modulated at about $25$~Hz to produce high average power and a consistent series of macropulses. 
It is worth noting that the pulse separation ($40$~ms) is somewhat smaller than the energy damping time of this electron beam (about $54$~ms).
This shows that reproducible production of pulsed \CM{2} OAM beams requires only modest damping.

In this Letter, we have reported the first experimental generation of OAM beams using an oscillator FEL by incorporating a specially designed spatial mask inside the FEL cavity.
Using masks of different dimensions, we have produced $\sim 458$~nm OAM beams in the four lowest orders, in the form of a coherent mixture of two LG modes of opposite helicities.
These OAM beams have been experimentally characterized to show an excellent mode quality factor, and consistent intensity and phase distributions compared with the theoretical ones.
Based on the $M^2$ measurements we have confirmed that the FEL OAM beams are dominated by the expected LG modes, possibly with a smaller contribution from a few other low-order modes.
The temporal structure of \CM{2} mode has been studied to show
a natural pulsing due to ac power modulation. 
Modulated by an external drive, this OAM beam produced using a $533$ MeV electron beam shows excellent reproducibility in terms of the pulse shape, pulse energy, and arrival time when the modulation frequency is less than $30$~Hz, with the best pulse energy consistency about $25$~Hz.
With this beam, a reasonably high intracavity laser power, on the order of $10$~W, has been realized, and even higher power ($\sim 100$~W) can be expected after further optimization of the optical mask and FEL mirrors.
We are working to extend the OAM generation to a range of wavelengths, from infrared to ultraviolet (UV) as allowed by the available FEL gain.

This work has demonstrated a novel method to generate OAM beams of various orders inside a laser cavity using the oscillator FEL.
The FEL technology opens the door for the generation and study of OAM beams with various features afforded by the FEL, including a wide spectral coverage, wavelength tunability, two-color lasing~\cite{YWu2015}, polarization manipulation and control~\cite{yan2019precision}, etc.
OAM FEL operation showcased in this Letter can be extended to higher photon energies, e.g. using
a future x-ray FEL oscillator~\cite{kim2008proposal}.
OAM FEL operation in EUV can be explored using either an oscillator~\cite{GStupakov2014}, or possibly, a regenerative amplifier~\cite{SBenson2010,YSocol2011}.
Furthermore, using the OAM FEL beam as the photon drive, a Compton light source can produce gamma-ray photons with OAM~\cite{jentschura2011generation,petrillo2016compton}.
If a Compton gamma-ray beam with sufficient flux can be generated, it will pave the way for a new class of nuclear physics experiments that can exploit novel physics phenomena associated with photon's OAM as a new degree of freedom in photo-nuclear reactions.

\begin{acknowledgments}
	We would like to thank the engineering and technical staff at DFELL/TUNL for their support.
	P.L., J.Y., H.H., S.F.M., V.G.P., Y.K.W. acknowledge DOE support under grant no. DE-FG02-97ER41033.
	A.A. acknowledges support from the U.S. Army Research Office grant no. W911NF-19-1-0022.  
	S.V.B. acknowledges DOE support under grant no. DE-AC05-06OR23177.
	The authors would like to thank Patrick Wallace for proofreading the manuscript.
\end{acknowledgments}


\begin{thebibliography}{71}%
\makeatletter
\providecommand \@ifxundefined [1]{%
 \@ifx{#1\undefined}
}%
\providecommand \@ifnum [1]{%
 \ifnum #1\expandafter \@firstoftwo
 \else \expandafter \@secondoftwo
 \fi
}%
\providecommand \@ifx [1]{%
 \ifx #1\expandafter \@firstoftwo
 \else \expandafter \@secondoftwo
 \fi
}%
\providecommand \natexlab [1]{#1}%
\providecommand \enquote  [1]{``#1''}%
\providecommand \bibnamefont  [1]{#1}%
\providecommand \bibfnamefont [1]{#1}%
\providecommand \citenamefont [1]{#1}%
\providecommand \href@noop [0]{\@secondoftwo}%
\providecommand \href [0]{\begingroup \@sanitize@url \@href}%
\providecommand \@href[1]{\@@startlink{#1}\@@href}%
\providecommand \@@href[1]{\endgroup#1\@@endlink}%
\providecommand \@sanitize@url [0]{\catcode `\\12\catcode `\$12\catcode
  `\&12\catcode `\#12\catcode `\^12\catcode `\_12\catcode `\%12\relax}%
\providecommand \@@startlink[1]{}%
\providecommand \@@endlink[0]{}%
\providecommand \url  [0]{\begingroup\@sanitize@url \@url }%
\providecommand \@url [1]{\endgroup\@href {#1}{\urlprefix }}%
\providecommand \urlprefix  [0]{URL }%
\providecommand \Eprint [0]{\href }%
\providecommand \doibase [0]{https://doi.org/}%
\providecommand \selectlanguage [0]{\@gobble}%
\providecommand \bibinfo  [0]{\@secondoftwo}%
\providecommand \bibfield  [0]{\@secondoftwo}%
\providecommand \translation [1]{[#1]}%
\providecommand \BibitemOpen [0]{}%
\providecommand \bibitemStop [0]{}%
\providecommand \bibitemNoStop [0]{.\EOS\space}%
\providecommand \EOS [0]{\spacefactor3000\relax}%
\providecommand \BibitemShut  [1]{\csname bibitem#1\endcsname}%
\let\auto@bib@innerbib\@empty
\bibitem [{\citenamefont {Allen}\ \emph {et~al.}(1992)\citenamefont {Allen},
  \citenamefont {Beijersbergen}, \citenamefont {Spreeuw},\ and\ \citenamefont
  {Woerdman}}]{allen1992orbital}%
  \BibitemOpen
  \bibfield  {author} {\bibinfo {author} {\bibfnamefont {L.}~\bibnamefont
  {Allen}}, \bibinfo {author} {\bibfnamefont {M.~W.}\ \bibnamefont
  {Beijersbergen}}, \bibinfo {author} {\bibfnamefont {R.~J.~C.}\ \bibnamefont
  {Spreeuw}},\ and\ \bibinfo {author} {\bibfnamefont {J.~P.}\ \bibnamefont
  {Woerdman}},\ }\bibfield  {title} {\bibinfo {title} {{Orbital angular
  momentum of light and the transformation of Laguerre-Gaussian laser modes}},\
  }\href@noop {} {\bibfield  {journal} {\bibinfo  {journal} {Phys. Rev. A}\
  }\textbf {\bibinfo {volume} {45}},\ \bibinfo {pages} {8185} (\bibinfo {year}
  {1992})}\BibitemShut {NoStop}%
\bibitem [{\citenamefont {Beijersbergen}\ \emph {et~al.}(1993)\citenamefont
  {Beijersbergen}, \citenamefont {Allen}, \citenamefont {van~der Veen},\ and\
  \citenamefont {Woerdman}}]{beijersbergen1993astigmatic}%
  \BibitemOpen
  \bibfield  {author} {\bibinfo {author} {\bibfnamefont {M.~W.}\ \bibnamefont
  {Beijersbergen}}, \bibinfo {author} {\bibfnamefont {L.}~\bibnamefont
  {Allen}}, \bibinfo {author} {\bibfnamefont {H.~E. L.~O.}\ \bibnamefont
  {van~der Veen}},\ and\ \bibinfo {author} {\bibfnamefont {J.~P.}\ \bibnamefont
  {Woerdman}},\ }\bibfield  {title} {\bibinfo {title} {Astigmatic laser mode
  converters and transfer of orbital angular momentum},\ }\href@noop {}
  {\bibfield  {journal} {\bibinfo  {journal} {Opt. Commun.}\ }\textbf {\bibinfo
  {volume} {96}},\ \bibinfo {pages} {123} (\bibinfo {year} {1993})}\BibitemShut
  {NoStop}%
\bibitem [{\citenamefont {Allen}\ \emph {et~al.}(1999)\citenamefont {Allen},
  \citenamefont {Padgett},\ and\ \citenamefont {Babiker}}]{allen1999iv}%
  \BibitemOpen
  \bibfield  {author} {\bibinfo {author} {\bibfnamefont {L.}~\bibnamefont
  {Allen}}, \bibinfo {author} {\bibfnamefont {M.~J.}\ \bibnamefont {Padgett}},\
  and\ \bibinfo {author} {\bibfnamefont {M.}~\bibnamefont {Babiker}},\
  }\bibfield  {title} {\bibinfo {title} {{IV The orbital angular momentum of
  light}},\ }in\ \href@noop {} {\emph {\bibinfo {booktitle} {Prog. Opt.}}},\
  Vol.~\bibinfo {volume} {39}\ (\bibinfo  {publisher} {Elsevier},\ \bibinfo
  {year} {1999})\ pp.\ \bibinfo {pages} {291--372}\BibitemShut {NoStop}%
\bibitem [{\citenamefont {He}\ \emph {et~al.}(1995)\citenamefont {He},
  \citenamefont {Heckenberg},\ and\ \citenamefont
  {Rubinsztein-Dunlop}}]{he1995optical}%
  \BibitemOpen
  \bibfield  {author} {\bibinfo {author} {\bibfnamefont {H.}~\bibnamefont
  {He}}, \bibinfo {author} {\bibfnamefont {N.~R.}\ \bibnamefont {Heckenberg}},\
  and\ \bibinfo {author} {\bibfnamefont {H.}~\bibnamefont
  {Rubinsztein-Dunlop}},\ }\bibfield  {title} {\bibinfo {title} {{Optical
  particle trapping with higher-order doughnut beams produced using high
  efficiency computer generated holograms}},\ }\href@noop {} {\bibfield
  {journal} {\bibinfo  {journal} {J. Mod. Opt.}\ }\textbf {\bibinfo {volume}
  {42}},\ \bibinfo {pages} {217} (\bibinfo {year} {1995})}\BibitemShut
  {NoStop}%
\bibitem [{\citenamefont {Beijersbergen}\ \emph {et~al.}(1994)\citenamefont
  {Beijersbergen}, \citenamefont {Coerwinkel}, \citenamefont {Kristensen},\
  and\ \citenamefont {Woerdman}}]{beijersbergen1994helical}%
  \BibitemOpen
  \bibfield  {author} {\bibinfo {author} {\bibfnamefont {M.~W.}\ \bibnamefont
  {Beijersbergen}}, \bibinfo {author} {\bibfnamefont {R.~P.~C.}\ \bibnamefont
  {Coerwinkel}}, \bibinfo {author} {\bibfnamefont {M.}~\bibnamefont
  {Kristensen}},\ and\ \bibinfo {author} {\bibfnamefont {J.~P.}\ \bibnamefont
  {Woerdman}},\ }\bibfield  {title} {\bibinfo {title} {Helical-wavefront laser
  beams produced with a spiral phaseplate},\ }\href@noop {} {\bibfield
  {journal} {\bibinfo  {journal} {Opt. Commun.}\ }\textbf {\bibinfo {volume}
  {112}},\ \bibinfo {pages} {321} (\bibinfo {year} {1994})}\BibitemShut
  {NoStop}%
\bibitem [{\citenamefont {Ito}\ \emph {et~al.}(2010)\citenamefont {Ito},
  \citenamefont {Kozawa},\ and\ \citenamefont {Sato}}]{ito2010generation}%
  \BibitemOpen
  \bibfield  {author} {\bibinfo {author} {\bibfnamefont {A.}~\bibnamefont
  {Ito}}, \bibinfo {author} {\bibfnamefont {Y.}~\bibnamefont {Kozawa}},\ and\
  \bibinfo {author} {\bibfnamefont {S.}~\bibnamefont {Sato}},\ }\bibfield
  {title} {\bibinfo {title} {{Generation of hollow scalar and vector beams
  using a spot-defect mirror}},\ }\href@noop {} {\bibfield  {journal} {\bibinfo
   {journal} {J. Opt. Soc. Am. A}\ }\textbf {\bibinfo {volume} {27}},\ \bibinfo
  {pages} {2072} (\bibinfo {year} {2010})}\BibitemShut {NoStop}%
\bibitem [{\citenamefont {Naidoo}\ \emph {et~al.}(2016)\citenamefont {Naidoo},
  \citenamefont {Roux}, \citenamefont {Dudley}, \citenamefont {Litvin},
  \citenamefont {Piccirillo}, \citenamefont {Marrucci},\ and\ \citenamefont
  {Forbes}}]{naidoo2016controlled}%
  \BibitemOpen
  \bibfield  {author} {\bibinfo {author} {\bibfnamefont {D.}~\bibnamefont
  {Naidoo}}, \bibinfo {author} {\bibfnamefont {F.~S.}\ \bibnamefont {Roux}},
  \bibinfo {author} {\bibfnamefont {A.}~\bibnamefont {Dudley}}, \bibinfo
  {author} {\bibfnamefont {I.}~\bibnamefont {Litvin}}, \bibinfo {author}
  {\bibfnamefont {B.}~\bibnamefont {Piccirillo}}, \bibinfo {author}
  {\bibfnamefont {L.}~\bibnamefont {Marrucci}},\ and\ \bibinfo {author}
  {\bibfnamefont {A.}~\bibnamefont {Forbes}},\ }\bibfield  {title} {\bibinfo
  {title} {Controlled generation of higher-order poincar{\'e} sphere beams from
  a laser},\ }\href@noop {} {\bibfield  {journal} {\bibinfo  {journal} {Nat.
  Photon.}\ }\textbf {\bibinfo {volume} {10}},\ \bibinfo {pages} {327}
  (\bibinfo {year} {2016})}\BibitemShut {NoStop}%
\bibitem [{\citenamefont {Franke-Arnold}\ \emph {et~al.}(2008)\citenamefont
  {Franke-Arnold}, \citenamefont {Allen},\ and\ \citenamefont
  {Padgett}}]{franke2008advances}%
  \BibitemOpen
  \bibfield  {author} {\bibinfo {author} {\bibfnamefont {S.}~\bibnamefont
  {Franke-Arnold}}, \bibinfo {author} {\bibfnamefont {L.}~\bibnamefont
  {Allen}},\ and\ \bibinfo {author} {\bibfnamefont {M.}~\bibnamefont
  {Padgett}},\ }\bibfield  {title} {\bibinfo {title} {Advances in optical
  angular momentum},\ }\href@noop {} {\bibfield  {journal} {\bibinfo  {journal}
  {Laser Photonics Rev.}\ }\textbf {\bibinfo {volume} {2}},\ \bibinfo {pages}
  {299} (\bibinfo {year} {2008})}\BibitemShut {NoStop}%
\bibitem [{\citenamefont {Yao}\ and\ \citenamefont
  {Padgett}(2011)}]{yao2011orbital}%
  \BibitemOpen
  \bibfield  {author} {\bibinfo {author} {\bibfnamefont {A.~M.}\ \bibnamefont
  {Yao}}\ and\ \bibinfo {author} {\bibfnamefont {M.~J.}\ \bibnamefont
  {Padgett}},\ }\bibfield  {title} {\bibinfo {title} {Orbital angular momentum:
  origins, behavior and applications},\ }\href@noop {} {\bibfield  {journal}
  {\bibinfo  {journal} {Adv. Opt. Photonics}\ }\textbf {\bibinfo {volume}
  {3}},\ \bibinfo {pages} {161} (\bibinfo {year} {2011})}\BibitemShut {NoStop}%
\bibitem [{\citenamefont {Rubinsztein-Dunlop}\ \emph
  {et~al.}(2016)\citenamefont {Rubinsztein-Dunlop} \emph
  {et~al.}}]{rubinsztein2016roadmap}%
  \BibitemOpen
  \bibfield  {author} {\bibinfo {author} {\bibfnamefont {H.}~\bibnamefont
  {Rubinsztein-Dunlop}} \emph {et~al.},\ }\bibfield  {title} {\bibinfo {title}
  {Roadmap on structured light},\ }\href@noop {} {\bibfield  {journal}
  {\bibinfo  {journal} {J. Opt.}\ }\textbf {\bibinfo {volume} {19}},\ \bibinfo
  {pages} {013001} (\bibinfo {year} {2016})}\BibitemShut {NoStop}%
\bibitem [{\citenamefont {Simpson}\ \emph {et~al.}(1996)\citenamefont
  {Simpson}, \citenamefont {Allen},\ and\ \citenamefont
  {Padgett}}]{simpson1996optical}%
  \BibitemOpen
  \bibfield  {author} {\bibinfo {author} {\bibfnamefont {N.~B.}\ \bibnamefont
  {Simpson}}, \bibinfo {author} {\bibfnamefont {L.}~\bibnamefont {Allen}},\
  and\ \bibinfo {author} {\bibfnamefont {M.~J.}\ \bibnamefont {Padgett}},\
  }\bibfield  {title} {\bibinfo {title} {Optical tweezers and optical spanners
  with {Laguerre--Gaussian} modes},\ }\href@noop {} {\bibfield  {journal}
  {\bibinfo  {journal} {J. Mod. Opt.}\ }\textbf {\bibinfo {volume} {43}},\
  \bibinfo {pages} {2485} (\bibinfo {year} {1996})}\BibitemShut {NoStop}%
\bibitem [{\citenamefont {Grier}(2003)}]{grier2003revolution}%
  \BibitemOpen
  \bibfield  {author} {\bibinfo {author} {\bibfnamefont {D.~G.}\ \bibnamefont
  {Grier}},\ }\bibfield  {title} {\bibinfo {title} {A revolution in optical
  manipulation},\ }\href@noop {} {\bibfield  {journal} {\bibinfo  {journal}
  {Nature}\ }\textbf {\bibinfo {volume} {424}},\ \bibinfo {pages} {810}
  (\bibinfo {year} {2003})}\BibitemShut {NoStop}%
\bibitem [{\citenamefont {Dasgupta}\ \emph {et~al.}(2011)\citenamefont
  {Dasgupta}, \citenamefont {Ahlawat}, \citenamefont {Verma},\ and\
  \citenamefont {Gupta}}]{dasgupta2011optical}%
  \BibitemOpen
  \bibfield  {author} {\bibinfo {author} {\bibfnamefont {R.}~\bibnamefont
  {Dasgupta}}, \bibinfo {author} {\bibfnamefont {S.}~\bibnamefont {Ahlawat}},
  \bibinfo {author} {\bibfnamefont {R.~S.}\ \bibnamefont {Verma}},\ and\
  \bibinfo {author} {\bibfnamefont {P.~K.}\ \bibnamefont {Gupta}},\ }\bibfield
  {title} {\bibinfo {title} {{Optical orientation and rotation of trapped red
  blood cells with Laguerre-Gaussian mode}},\ }\href@noop {} {\bibfield
  {journal} {\bibinfo  {journal} {Opt. Express}\ }\textbf {\bibinfo {volume}
  {19}},\ \bibinfo {pages} {7680} (\bibinfo {year} {2011})}\BibitemShut
  {NoStop}%
\bibitem [{\citenamefont {Kuga}\ \emph {et~al.}(1997)\citenamefont {Kuga},
  \citenamefont {Torii}, \citenamefont {Shiokawa}, \citenamefont {Hirano},
  \citenamefont {Shimizu},\ and\ \citenamefont {Sasada}}]{kuga1997novel}%
  \BibitemOpen
  \bibfield  {author} {\bibinfo {author} {\bibfnamefont {T.}~\bibnamefont
  {Kuga}}, \bibinfo {author} {\bibfnamefont {Y.}~\bibnamefont {Torii}},
  \bibinfo {author} {\bibfnamefont {N.}~\bibnamefont {Shiokawa}}, \bibinfo
  {author} {\bibfnamefont {T.}~\bibnamefont {Hirano}}, \bibinfo {author}
  {\bibfnamefont {Y.}~\bibnamefont {Shimizu}},\ and\ \bibinfo {author}
  {\bibfnamefont {H.}~\bibnamefont {Sasada}},\ }\bibfield  {title} {\bibinfo
  {title} {Novel optical trap of atoms with a doughnut beam},\ }\href@noop {}
  {\bibfield  {journal} {\bibinfo  {journal} {Phys. Rev. Lett.}\ }\textbf
  {\bibinfo {volume} {78}},\ \bibinfo {pages} {4713} (\bibinfo {year}
  {1997})}\BibitemShut {NoStop}%
\bibitem [{\citenamefont {Tabosa}\ and\ \citenamefont
  {Petrov}(1999)}]{tabosa1999optical}%
  \BibitemOpen
  \bibfield  {author} {\bibinfo {author} {\bibfnamefont {J.}~\bibnamefont
  {Tabosa}}\ and\ \bibinfo {author} {\bibfnamefont {D.}~\bibnamefont
  {Petrov}},\ }\bibfield  {title} {\bibinfo {title} {Optical pumping of orbital
  angular momentum of light in cold cesium atoms},\ }\href@noop {} {\bibfield
  {journal} {\bibinfo  {journal} {Phys. Rev. Lett.}\ }\textbf {\bibinfo
  {volume} {83}},\ \bibinfo {pages} {4967} (\bibinfo {year}
  {1999})}\BibitemShut {NoStop}%
\bibitem [{\citenamefont {Andersen}\ \emph {et~al.}(2006)\citenamefont
  {Andersen}, \citenamefont {Ryu}, \citenamefont {Clad{\'e}}, \citenamefont
  {Natarajan}, \citenamefont {Vaziri}, \citenamefont {Helmerson},\ and\
  \citenamefont {Phillips}}]{andersen2006quantized}%
  \BibitemOpen
  \bibfield  {author} {\bibinfo {author} {\bibfnamefont {M.}~\bibnamefont
  {Andersen}}, \bibinfo {author} {\bibfnamefont {C.}~\bibnamefont {Ryu}},
  \bibinfo {author} {\bibfnamefont {P.}~\bibnamefont {Clad{\'e}}}, \bibinfo
  {author} {\bibfnamefont {V.}~\bibnamefont {Natarajan}}, \bibinfo {author}
  {\bibfnamefont {A.}~\bibnamefont {Vaziri}}, \bibinfo {author} {\bibfnamefont
  {K.}~\bibnamefont {Helmerson}},\ and\ \bibinfo {author} {\bibfnamefont
  {W.~D.}\ \bibnamefont {Phillips}},\ }\bibfield  {title} {\bibinfo {title}
  {Quantized rotation of atoms from photons with orbital angular momentum},\
  }\href@noop {} {\bibfield  {journal} {\bibinfo  {journal} {Phys. Rev. Lett.}\
  }\textbf {\bibinfo {volume} {97}},\ \bibinfo {pages} {170406} (\bibinfo
  {year} {2006})}\BibitemShut {NoStop}%
\bibitem [{\citenamefont {Wright}\ \emph {et~al.}(2008)\citenamefont {Wright},
  \citenamefont {Leslie},\ and\ \citenamefont {Bigelow}}]{wright2008optical}%
  \BibitemOpen
  \bibfield  {author} {\bibinfo {author} {\bibfnamefont {K.~C.}\ \bibnamefont
  {Wright}}, \bibinfo {author} {\bibfnamefont {L.~S.}\ \bibnamefont {Leslie}},\
  and\ \bibinfo {author} {\bibfnamefont {N.~P.}\ \bibnamefont {Bigelow}},\
  }\bibfield  {title} {\bibinfo {title} {Optical control of the internal and
  external angular momentum of a {Bose-Einstein} condensate},\ }\href@noop {}
  {\bibfield  {journal} {\bibinfo  {journal} {Phys. Rev. A}\ }\textbf {\bibinfo
  {volume} {77}},\ \bibinfo {pages} {041601} (\bibinfo {year}
  {2008})}\BibitemShut {NoStop}%
\bibitem [{\citenamefont {Molina-Terriza}\ \emph {et~al.}(2001)\citenamefont
  {Molina-Terriza}, \citenamefont {Torres},\ and\ \citenamefont
  {Torner}}]{molina2001management}%
  \BibitemOpen
  \bibfield  {author} {\bibinfo {author} {\bibfnamefont {G.}~\bibnamefont
  {Molina-Terriza}}, \bibinfo {author} {\bibfnamefont {J.~P.}\ \bibnamefont
  {Torres}},\ and\ \bibinfo {author} {\bibfnamefont {L.}~\bibnamefont
  {Torner}},\ }\bibfield  {title} {\bibinfo {title} {Management of the angular
  momentum of light: preparation of photons in multidimensional vector states
  of angular momentum},\ }\href@noop {} {\bibfield  {journal} {\bibinfo
  {journal} {Phys. Rev. Lett.}\ }\textbf {\bibinfo {volume} {88}},\ \bibinfo
  {pages} {013601} (\bibinfo {year} {2001})}\BibitemShut {NoStop}%
\bibitem [{\citenamefont {Gibson}\ \emph {et~al.}(2004)\citenamefont {Gibson},
  \citenamefont {Courtial}, \citenamefont {Padgett}, \citenamefont {Vasnetsov},
  \citenamefont {Pas’ko}, \citenamefont {Barnett},\ and\ \citenamefont
  {Franke-Arnold}}]{gibson2004free}%
  \BibitemOpen
  \bibfield  {author} {\bibinfo {author} {\bibfnamefont {G.}~\bibnamefont
  {Gibson}}, \bibinfo {author} {\bibfnamefont {J.}~\bibnamefont {Courtial}},
  \bibinfo {author} {\bibfnamefont {M.~J.}\ \bibnamefont {Padgett}}, \bibinfo
  {author} {\bibfnamefont {M.}~\bibnamefont {Vasnetsov}}, \bibinfo {author}
  {\bibfnamefont {V.}~\bibnamefont {Pas’ko}}, \bibinfo {author}
  {\bibfnamefont {S.~M.}\ \bibnamefont {Barnett}},\ and\ \bibinfo {author}
  {\bibfnamefont {S.}~\bibnamefont {Franke-Arnold}},\ }\bibfield  {title}
  {\bibinfo {title} {Free-space information transfer using light beams carrying
  orbital angular momentum},\ }\href@noop {} {\bibfield  {journal} {\bibinfo
  {journal} {Opt. Express}\ }\textbf {\bibinfo {volume} {12}},\ \bibinfo
  {pages} {5448} (\bibinfo {year} {2004})}\BibitemShut {NoStop}%
\bibitem [{\citenamefont {Nagali}\ \emph {et~al.}(2009)\citenamefont {Nagali},
  \citenamefont {Sciarrino}, \citenamefont {De~Martini}, \citenamefont
  {Marrucci}, \citenamefont {Piccirillo}, \citenamefont {Karimi},\ and\
  \citenamefont {Santamato}}]{nagali2009quantum}%
  \BibitemOpen
  \bibfield  {author} {\bibinfo {author} {\bibfnamefont {E.}~\bibnamefont
  {Nagali}}, \bibinfo {author} {\bibfnamefont {F.}~\bibnamefont {Sciarrino}},
  \bibinfo {author} {\bibfnamefont {F.}~\bibnamefont {De~Martini}}, \bibinfo
  {author} {\bibfnamefont {L.}~\bibnamefont {Marrucci}}, \bibinfo {author}
  {\bibfnamefont {B.}~\bibnamefont {Piccirillo}}, \bibinfo {author}
  {\bibfnamefont {E.}~\bibnamefont {Karimi}},\ and\ \bibinfo {author}
  {\bibfnamefont {E.}~\bibnamefont {Santamato}},\ }\bibfield  {title} {\bibinfo
  {title} {Quantum information transfer from spin to orbital angular momentum
  of photons},\ }\href@noop {} {\bibfield  {journal} {\bibinfo  {journal}
  {Phys. Rev. Lett.}\ }\textbf {\bibinfo {volume} {103}},\ \bibinfo {pages}
  {013601} (\bibinfo {year} {2009})}\BibitemShut {NoStop}%
\bibitem [{\citenamefont {Vallone}\ \emph {et~al.}(2014)\citenamefont
  {Vallone}, \citenamefont {D’Ambrosio}, \citenamefont {Sponselli},
  \citenamefont {Slussarenko}, \citenamefont {Marrucci}, \citenamefont
  {Sciarrino},\ and\ \citenamefont {Villoresi}}]{vallone2014free}%
  \BibitemOpen
  \bibfield  {author} {\bibinfo {author} {\bibfnamefont {G.}~\bibnamefont
  {Vallone}}, \bibinfo {author} {\bibfnamefont {V.}~\bibnamefont
  {D’Ambrosio}}, \bibinfo {author} {\bibfnamefont {A.}~\bibnamefont
  {Sponselli}}, \bibinfo {author} {\bibfnamefont {S.}~\bibnamefont
  {Slussarenko}}, \bibinfo {author} {\bibfnamefont {L.}~\bibnamefont
  {Marrucci}}, \bibinfo {author} {\bibfnamefont {F.}~\bibnamefont
  {Sciarrino}},\ and\ \bibinfo {author} {\bibfnamefont {P.}~\bibnamefont
  {Villoresi}},\ }\bibfield  {title} {\bibinfo {title} {Free-space quantum key
  distribution by rotation-invariant twisted photons},\ }\href@noop {}
  {\bibfield  {journal} {\bibinfo  {journal} {Phys. Rev. Lett.}\ }\textbf
  {\bibinfo {volume} {113}},\ \bibinfo {pages} {060503} (\bibinfo {year}
  {2014})}\BibitemShut {NoStop}%
\bibitem [{\citenamefont {Schmiegelow}\ \emph {et~al.}(2016)\citenamefont
  {Schmiegelow}, \citenamefont {Schulz}, \citenamefont {Kaufmann},
  \citenamefont {Ruster}, \citenamefont {Poschinger},\ and\ \citenamefont
  {Schmidt-Kaler}}]{schmiegelow2016transfer}%
  \BibitemOpen
  \bibfield  {author} {\bibinfo {author} {\bibfnamefont {C.~T.}\ \bibnamefont
  {Schmiegelow}}, \bibinfo {author} {\bibfnamefont {J.}~\bibnamefont {Schulz}},
  \bibinfo {author} {\bibfnamefont {H.}~\bibnamefont {Kaufmann}}, \bibinfo
  {author} {\bibfnamefont {T.}~\bibnamefont {Ruster}}, \bibinfo {author}
  {\bibfnamefont {U.~G.}\ \bibnamefont {Poschinger}},\ and\ \bibinfo {author}
  {\bibfnamefont {F.}~\bibnamefont {Schmidt-Kaler}},\ }\bibfield  {title}
  {\bibinfo {title} {{Transfer of optical orbital angular momentum to a bound
  electron}},\ }\href@noop {} {\bibfield  {journal} {\bibinfo  {journal} {Nat.
  Commun.}\ }\textbf {\bibinfo {volume} {7}},\ \bibinfo {pages} {12998}
  (\bibinfo {year} {2016})}\BibitemShut {NoStop}%
\bibitem [{\citenamefont {Afanasev}\ \emph
  {et~al.}(2018{\natexlab{a}})\citenamefont {Afanasev}, \citenamefont
  {Carlson}, \citenamefont {Schmiegelow}, \citenamefont {Schulz}, \citenamefont
  {Schmidt-Kaler},\ and\ \citenamefont {Solyanik}}]{afanasev2018experimental}%
  \BibitemOpen
  \bibfield  {author} {\bibinfo {author} {\bibfnamefont {A.}~\bibnamefont
  {Afanasev}}, \bibinfo {author} {\bibfnamefont {C.~E.}\ \bibnamefont
  {Carlson}}, \bibinfo {author} {\bibfnamefont {C.~T.}\ \bibnamefont
  {Schmiegelow}}, \bibinfo {author} {\bibfnamefont {J.}~\bibnamefont {Schulz}},
  \bibinfo {author} {\bibfnamefont {F.}~\bibnamefont {Schmidt-Kaler}},\ and\
  \bibinfo {author} {\bibfnamefont {M.}~\bibnamefont {Solyanik}},\ }\bibfield
  {title} {\bibinfo {title} {Experimental verification of position-dependent
  angular-momentum selection rules for absorption of twisted light by a bound
  electron},\ }\href@noop {} {\bibfield  {journal} {\bibinfo  {journal} {New J.
  Phys.}\ }\textbf {\bibinfo {volume} {20}},\ \bibinfo {pages} {023032}
  (\bibinfo {year} {2018}{\natexlab{a}})}\BibitemShut {NoStop}%
\bibitem [{\citenamefont {Pic{\'o}n}\ \emph {et~al.}(2010)\citenamefont
  {Pic{\'o}n}, \citenamefont {Mompart}, \citenamefont {de~Aldana},
  \citenamefont {Plaja}, \citenamefont {Calvo},\ and\ \citenamefont
  {Roso}}]{picon2010photoionization}%
  \BibitemOpen
  \bibfield  {author} {\bibinfo {author} {\bibfnamefont {A.}~\bibnamefont
  {Pic{\'o}n}}, \bibinfo {author} {\bibfnamefont {J.}~\bibnamefont {Mompart}},
  \bibinfo {author} {\bibfnamefont {J.~R.~V.}\ \bibnamefont {de~Aldana}},
  \bibinfo {author} {\bibfnamefont {L.}~\bibnamefont {Plaja}}, \bibinfo
  {author} {\bibfnamefont {G.~F.}\ \bibnamefont {Calvo}},\ and\ \bibinfo
  {author} {\bibfnamefont {L.}~\bibnamefont {Roso}},\ }\bibfield  {title}
  {\bibinfo {title} {Photoionization with orbital angular momentum beams},\
  }\href@noop {} {\bibfield  {journal} {\bibinfo  {journal} {Opt. Express}\
  }\textbf {\bibinfo {volume} {18}},\ \bibinfo {pages} {3660} (\bibinfo {year}
  {2010})}\BibitemShut {NoStop}%
\bibitem [{\citenamefont {Sakdinawat}\ and\ \citenamefont
  {Liu}(2007)}]{sakdinawat2007soft}%
  \BibitemOpen
  \bibfield  {author} {\bibinfo {author} {\bibfnamefont {A.}~\bibnamefont
  {Sakdinawat}}\ and\ \bibinfo {author} {\bibfnamefont {Y.}~\bibnamefont
  {Liu}},\ }\bibfield  {title} {\bibinfo {title} {Soft-x-ray microscopy using
  spiral zone plates},\ }\href@noop {} {\bibfield  {journal} {\bibinfo
  {journal} {Opt. Lett.}\ }\textbf {\bibinfo {volume} {32}},\ \bibinfo {pages}
  {2635} (\bibinfo {year} {2007})}\BibitemShut {NoStop}%
\bibitem [{\citenamefont {van Veenendaal}\ and\ \citenamefont
  {McNulty}(2007)}]{van2007prediction}%
  \BibitemOpen
  \bibfield  {author} {\bibinfo {author} {\bibfnamefont {M.}~\bibnamefont {van
  Veenendaal}}\ and\ \bibinfo {author} {\bibfnamefont {I.}~\bibnamefont
  {McNulty}},\ }\bibfield  {title} {\bibinfo {title} {{Prediction of strong
  dichroism induced by x rays carrying orbital momentum}},\ }\href@noop {}
  {\bibfield  {journal} {\bibinfo  {journal} {Phys. Rev. Lett.}\ }\textbf
  {\bibinfo {volume} {98}},\ \bibinfo {pages} {157401} (\bibinfo {year}
  {2007})}\BibitemShut {NoStop}%
\bibitem [{\citenamefont {W{\"a}tzel}\ \emph {et~al.}(2016)\citenamefont
  {W{\"a}tzel}, \citenamefont {Pavlyukh}, \citenamefont {Sch{\"a}ffer},\ and\
  \citenamefont {Berakdar}}]{watzel2016optical}%
  \BibitemOpen
  \bibfield  {author} {\bibinfo {author} {\bibfnamefont {J.}~\bibnamefont
  {W{\"a}tzel}}, \bibinfo {author} {\bibfnamefont {Y.}~\bibnamefont
  {Pavlyukh}}, \bibinfo {author} {\bibfnamefont {A.}~\bibnamefont
  {Sch{\"a}ffer}},\ and\ \bibinfo {author} {\bibfnamefont {J.}~\bibnamefont
  {Berakdar}},\ }\bibfield  {title} {\bibinfo {title} {Optical vortex driven
  charge current loop and optomagnetism in fullerenes},\ }\href@noop {}
  {\bibfield  {journal} {\bibinfo  {journal} {Carbon}\ }\textbf {\bibinfo
  {volume} {99}},\ \bibinfo {pages} {439} (\bibinfo {year} {2016})}\BibitemShut
  {NoStop}%
\bibitem [{\citenamefont {Fujita}\ and\ \citenamefont
  {Sato}(2017)}]{fujita2017ultrafast}%
  \BibitemOpen
  \bibfield  {author} {\bibinfo {author} {\bibfnamefont {H.}~\bibnamefont
  {Fujita}}\ and\ \bibinfo {author} {\bibfnamefont {M.}~\bibnamefont {Sato}},\
  }\bibfield  {title} {\bibinfo {title} {Ultrafast generation of skyrmionic
  defects with vortex beams: Printing laser profiles on magnets},\ }\href@noop
  {} {\bibfield  {journal} {\bibinfo  {journal} {Phys. Rev. B}\ }\textbf
  {\bibinfo {volume} {95}},\ \bibinfo {pages} {054421} (\bibinfo {year}
  {2017})}\BibitemShut {NoStop}%
\bibitem [{\citenamefont {Afanasev}\ \emph
  {et~al.}(2018{\natexlab{b}})\citenamefont {Afanasev}, \citenamefont {Serbo},\
  and\ \citenamefont {Solyanik}}]{afanasev2018radiative}%
  \BibitemOpen
  \bibfield  {author} {\bibinfo {author} {\bibfnamefont {A.}~\bibnamefont
  {Afanasev}}, \bibinfo {author} {\bibfnamefont {V.~G.}\ \bibnamefont
  {Serbo}},\ and\ \bibinfo {author} {\bibfnamefont {M.}~\bibnamefont
  {Solyanik}},\ }\bibfield  {title} {\bibinfo {title} {Radiative capture of
  cold neutrons by protons and deuteron photodisintegration with twisted
  beams},\ }\href@noop {} {\bibfield  {journal} {\bibinfo  {journal} {J. Phys.
  G Nucl. Partic. Phys.}\ }\textbf {\bibinfo {volume} {45}},\ \bibinfo {pages}
  {055102} (\bibinfo {year} {2018}{\natexlab{b}})}\BibitemShut {NoStop}%
\bibitem [{\citenamefont {Afanasev}\ \emph {et~al.}(2017)\citenamefont
  {Afanasev}, \citenamefont {Carlson},\ and\ \citenamefont
  {Solyanik}}]{afanasev2017circular}%
  \BibitemOpen
  \bibfield  {author} {\bibinfo {author} {\bibfnamefont {A.}~\bibnamefont
  {Afanasev}}, \bibinfo {author} {\bibfnamefont {C.~E.}\ \bibnamefont
  {Carlson}},\ and\ \bibinfo {author} {\bibfnamefont {M.}~\bibnamefont
  {Solyanik}},\ }\bibfield  {title} {\bibinfo {title} {Circular dichroism of
  twisted photons in non-chiral atomic matter},\ }\href@noop {} {\bibfield
  {journal} {\bibinfo  {journal} {J. Opt.}\ }\textbf {\bibinfo {volume} {19}},\
  \bibinfo {pages} {105401} (\bibinfo {year} {2017})}\BibitemShut {NoStop}%
\bibitem [{\citenamefont {Ivanov}\ \emph
  {et~al.}(2020{\natexlab{a}})\citenamefont {Ivanov}, \citenamefont
  {Korchagin}, \citenamefont {Pimikov},\ and\ \citenamefont
  {Zhang}}]{ivanov2020doing}%
  \BibitemOpen
  \bibfield  {author} {\bibinfo {author} {\bibfnamefont {I.~P.}\ \bibnamefont
  {Ivanov}}, \bibinfo {author} {\bibfnamefont {N.}~\bibnamefont {Korchagin}},
  \bibinfo {author} {\bibfnamefont {A.}~\bibnamefont {Pimikov}},\ and\ \bibinfo
  {author} {\bibfnamefont {P.}~\bibnamefont {Zhang}},\ }\bibfield  {title}
  {\bibinfo {title} {Doing spin physics with unpolarized particles},\
  }\href@noop {} {\bibfield  {journal} {\bibinfo  {journal} {Phys. Rev. Lett.}\
  }\textbf {\bibinfo {volume} {124}},\ \bibinfo {pages} {192001} (\bibinfo
  {year} {2020}{\natexlab{a}})}\BibitemShut {NoStop}%
\bibitem [{\citenamefont {Ivanov}\ \emph
  {et~al.}(2020{\natexlab{b}})\citenamefont {Ivanov}, \citenamefont
  {Korchagin}, \citenamefont {Pimikov},\ and\ \citenamefont
  {Zhang}}]{ivanov2020kinematic}%
  \BibitemOpen
  \bibfield  {author} {\bibinfo {author} {\bibfnamefont {I.~P.}\ \bibnamefont
  {Ivanov}}, \bibinfo {author} {\bibfnamefont {N.}~\bibnamefont {Korchagin}},
  \bibinfo {author} {\bibfnamefont {A.}~\bibnamefont {Pimikov}},\ and\ \bibinfo
  {author} {\bibfnamefont {P.}~\bibnamefont {Zhang}},\ }\bibfield  {title}
  {\bibinfo {title} {Kinematic surprises in twisted-particle collisions},\
  }\href@noop {} {\bibfield  {journal} {\bibinfo  {journal} {Phys. Rev. D}\
  }\textbf {\bibinfo {volume} {101}},\ \bibinfo {pages} {016007} (\bibinfo
  {year} {2020}{\natexlab{b}})}\BibitemShut {NoStop}%
\bibitem [{\citenamefont {Afanasev}\ \emph
  {et~al.}(2018{\natexlab{c}})\citenamefont {Afanasev}, \citenamefont
  {Carlson},\ and\ \citenamefont {Solyanik}}]{afanasev2018atomic}%
  \BibitemOpen
  \bibfield  {author} {\bibinfo {author} {\bibfnamefont {A.}~\bibnamefont
  {Afanasev}}, \bibinfo {author} {\bibfnamefont {C.~E.}\ \bibnamefont
  {Carlson}},\ and\ \bibinfo {author} {\bibfnamefont {M.}~\bibnamefont
  {Solyanik}},\ }\bibfield  {title} {\bibinfo {title} {{Atomic spectroscopy
  with twisted photons: Separation of M1- E2 mixed multipoles}},\ }\href@noop
  {} {\bibfield  {journal} {\bibinfo  {journal} {Phys. Rev. A}\ }\textbf
  {\bibinfo {volume} {97}},\ \bibinfo {pages} {023422} (\bibinfo {year}
  {2018}{\natexlab{c}})}\BibitemShut {NoStop}%
\bibitem [{\citenamefont {Taira}\ and\ \citenamefont
  {Katoh}(2018)}]{taira2018generation}%
  \BibitemOpen
  \bibfield  {author} {\bibinfo {author} {\bibfnamefont {Y.}~\bibnamefont
  {Taira}}\ and\ \bibinfo {author} {\bibfnamefont {M.}~\bibnamefont {Katoh}},\
  }\bibfield  {title} {\bibinfo {title} {{Generation of optical vortices by
  nonlinear inverse Thomson scattering at arbitrary angle interactions}},\
  }\href@noop {} {\bibfield  {journal} {\bibinfo  {journal} {Astrophys. J.}\
  }\textbf {\bibinfo {volume} {860}},\ \bibinfo {pages} {45} (\bibinfo {year}
  {2018})}\BibitemShut {NoStop}%
\bibitem [{\citenamefont {Katoh}\ \emph {et~al.}(2017)\citenamefont {Katoh},
  \citenamefont {Fujimoto}, \citenamefont {Kawaguchi}, \citenamefont
  {Tsuchiya}, \citenamefont {Ohmi}, \citenamefont {Kaneyasu}, \citenamefont
  {Taira}, \citenamefont {Hosaka}, \citenamefont {Mochihashi},\ and\
  \citenamefont {Takashima}}]{katoh2017angular}%
  \BibitemOpen
  \bibfield  {author} {\bibinfo {author} {\bibfnamefont {M.}~\bibnamefont
  {Katoh}}, \bibinfo {author} {\bibfnamefont {M.}~\bibnamefont {Fujimoto}},
  \bibinfo {author} {\bibfnamefont {H.}~\bibnamefont {Kawaguchi}}, \bibinfo
  {author} {\bibfnamefont {K.}~\bibnamefont {Tsuchiya}}, \bibinfo {author}
  {\bibfnamefont {K.}~\bibnamefont {Ohmi}}, \bibinfo {author} {\bibfnamefont
  {T.}~\bibnamefont {Kaneyasu}}, \bibinfo {author} {\bibfnamefont
  {Y.}~\bibnamefont {Taira}}, \bibinfo {author} {\bibfnamefont
  {M.}~\bibnamefont {Hosaka}}, \bibinfo {author} {\bibfnamefont
  {A.}~\bibnamefont {Mochihashi}},\ and\ \bibinfo {author} {\bibfnamefont
  {Y.}~\bibnamefont {Takashima}},\ }\bibfield  {title} {\bibinfo {title}
  {Angular momentum of twisted radiation from an electron in spiral motion},\
  }\href@noop {} {\bibfield  {journal} {\bibinfo  {journal} {Phys. Rev. Lett.}\
  }\textbf {\bibinfo {volume} {118}},\ \bibinfo {pages} {094801} (\bibinfo
  {year} {2017})}\BibitemShut {NoStop}%
\bibitem [{\citenamefont {Maruyama}\ \emph
  {et~al.}(2019{\natexlab{a}})\citenamefont {Maruyama}, \citenamefont
  {Hayakawa},\ and\ \citenamefont {Kajino}}]{maruyama2019compton}%
  \BibitemOpen
  \bibfield  {author} {\bibinfo {author} {\bibfnamefont {T.}~\bibnamefont
  {Maruyama}}, \bibinfo {author} {\bibfnamefont {T.}~\bibnamefont {Hayakawa}},\
  and\ \bibinfo {author} {\bibfnamefont {T.}~\bibnamefont {Kajino}},\
  }\bibfield  {title} {\bibinfo {title} {{Compton Scattering of $\gamma$-Ray
  Vortex with Laguerre Gaussian Wave Function}},\ }\href@noop {} {\bibfield
  {journal} {\bibinfo  {journal} {Sci. Rep.}\ }\textbf {\bibinfo {volume}
  {9}},\ \bibinfo {pages} {51} (\bibinfo {year}
  {2019}{\natexlab{a}})}\BibitemShut {NoStop}%
\bibitem [{\citenamefont {Maruyama}\ \emph
  {et~al.}(2019{\natexlab{b}})\citenamefont {Maruyama}, \citenamefont
  {Hayakawa},\ and\ \citenamefont {Kajino}}]{maruyama2019comptonHG}%
  \BibitemOpen
  \bibfield  {author} {\bibinfo {author} {\bibfnamefont {T.}~\bibnamefont
  {Maruyama}}, \bibinfo {author} {\bibfnamefont {T.}~\bibnamefont {Hayakawa}},\
  and\ \bibinfo {author} {\bibfnamefont {T.}~\bibnamefont {Kajino}},\
  }\bibfield  {title} {\bibinfo {title} {{Compton Scattering of Hermite
  Gaussian Wave $\gamma$ Ray}},\ }\href@noop {} {\bibfield  {journal} {\bibinfo
   {journal} {Sci. Rep.}\ }\textbf {\bibinfo {volume} {9}},\ \bibinfo {pages}
  {7998} (\bibinfo {year} {2019}{\natexlab{b}})}\BibitemShut {NoStop}%
\bibitem [{\citenamefont {Afanasev}\ \emph {et~al.}(2020)\citenamefont
  {Afanasev}, \citenamefont {Carlson},\ and\ \citenamefont
  {Mukherjee}}]{afanasev2020recoil}%
  \BibitemOpen
  \bibfield  {author} {\bibinfo {author} {\bibfnamefont {A.}~\bibnamefont
  {Afanasev}}, \bibinfo {author} {\bibfnamefont {C.~E.}\ \bibnamefont
  {Carlson}},\ and\ \bibinfo {author} {\bibfnamefont {A.}~\bibnamefont
  {Mukherjee}},\ }\bibfield  {title} {\bibinfo {title} {{Recoil Momentum
  Effects in Quantum Processes Induced by Twisted Photons}},\ }\href@noop {}
  {\bibfield  {journal} {\bibinfo  {journal} {arXiv preprint arXiv:2007.05816}\
  } (\bibinfo {year} {2020})}\BibitemShut {NoStop}%
\bibitem [{\citenamefont {Sasaki}\ and\ \citenamefont
  {McNulty}(2008)}]{SSasaki2008}%
  \BibitemOpen
  \bibfield  {author} {\bibinfo {author} {\bibfnamefont {S.}~\bibnamefont
  {Sasaki}}\ and\ \bibinfo {author} {\bibfnamefont {I.}~\bibnamefont
  {McNulty}},\ }\bibfield  {title} {\bibinfo {title} {Proposal for generating
  brilliant x-ray beams carrying orbital angular momentum},\ }\href@noop {}
  {\bibfield  {journal} {\bibinfo  {journal} {Phys. Rev. Lett.}\ }\textbf
  {\bibinfo {volume} {100}},\ \bibinfo {pages} {124801} (\bibinfo {year}
  {2008})}\BibitemShut {NoStop}%
\bibitem [{\citenamefont {Afanasev}\ and\ \citenamefont
  {Mikhailichenko}(2011)}]{afanasev2011generation}%
  \BibitemOpen
  \bibfield  {author} {\bibinfo {author} {\bibfnamefont {A.}~\bibnamefont
  {Afanasev}}\ and\ \bibinfo {author} {\bibfnamefont {A.}~\bibnamefont
  {Mikhailichenko}},\ }\bibfield  {title} {\bibinfo {title} {On generation of
  photons carrying orbital angular momentum in the helical undulator},\
  }\href@noop {} {\bibfield  {journal} {\bibinfo  {journal} {arXiv preprint
  arXiv:1109.1603}\ } (\bibinfo {year} {2011})}\BibitemShut {NoStop}%
\bibitem [{\citenamefont {Bahrdt}\ \emph {et~al.}(2013)\citenamefont {Bahrdt},
  \citenamefont {Holldack}, \citenamefont {Kuske}, \citenamefont {M{\"u}ller},
  \citenamefont {Scheer},\ and\ \citenamefont {Schmid}}]{JBahrdt2013}%
  \BibitemOpen
  \bibfield  {author} {\bibinfo {author} {\bibfnamefont {J.}~\bibnamefont
  {Bahrdt}}, \bibinfo {author} {\bibfnamefont {K.}~\bibnamefont {Holldack}},
  \bibinfo {author} {\bibfnamefont {P.}~\bibnamefont {Kuske}}, \bibinfo
  {author} {\bibfnamefont {R.}~\bibnamefont {M{\"u}ller}}, \bibinfo {author}
  {\bibfnamefont {M.}~\bibnamefont {Scheer}},\ and\ \bibinfo {author}
  {\bibfnamefont {P.}~\bibnamefont {Schmid}},\ }\bibfield  {title} {\bibinfo
  {title} {First observation of photons carrying orbital angular momentum in
  undulator radiation},\ }\href@noop {} {\bibfield  {journal} {\bibinfo
  {journal} {Phys. Rev. Lett.}\ }\textbf {\bibinfo {volume} {111}},\ \bibinfo
  {pages} {034801} (\bibinfo {year} {2013})}\BibitemShut {NoStop}%
\bibitem [{\citenamefont {Hemsing}\ \emph {et~al.}(2013)\citenamefont
  {Hemsing}, \citenamefont {Knyazik}, \citenamefont {Dunning}, \citenamefont
  {Xiang}, \citenamefont {Marinelli}, \citenamefont {Hast},\ and\ \citenamefont
  {Rosenzweig}}]{EHemsing2013}%
  \BibitemOpen
  \bibfield  {author} {\bibinfo {author} {\bibfnamefont {E.}~\bibnamefont
  {Hemsing}}, \bibinfo {author} {\bibfnamefont {A.}~\bibnamefont {Knyazik}},
  \bibinfo {author} {\bibfnamefont {M.}~\bibnamefont {Dunning}}, \bibinfo
  {author} {\bibfnamefont {D.}~\bibnamefont {Xiang}}, \bibinfo {author}
  {\bibfnamefont {A.}~\bibnamefont {Marinelli}}, \bibinfo {author}
  {\bibfnamefont {C.}~\bibnamefont {Hast}},\ and\ \bibinfo {author}
  {\bibfnamefont {J.~B.}\ \bibnamefont {Rosenzweig}},\ }\bibfield  {title}
  {\bibinfo {title} {Coherent optical vortices from relativistic electron
  beams},\ }\href@noop {} {\bibfield  {journal} {\bibinfo  {journal} {Nat.
  Phys.}\ }\textbf {\bibinfo {volume} {9}},\ \bibinfo {pages} {549} (\bibinfo
  {year} {2013})}\BibitemShut {NoStop}%
\bibitem [{\citenamefont {Hemsing}\ \emph {et~al.}(2014)\citenamefont
  {Hemsing}, \citenamefont {Dunning}, \citenamefont {Hast}, \citenamefont
  {Raubenheimer},\ and\ \citenamefont {Xiang}}]{EHemsing2014}%
  \BibitemOpen
  \bibfield  {author} {\bibinfo {author} {\bibfnamefont {E.}~\bibnamefont
  {Hemsing}}, \bibinfo {author} {\bibfnamefont {M.}~\bibnamefont {Dunning}},
  \bibinfo {author} {\bibfnamefont {C.}~\bibnamefont {Hast}}, \bibinfo {author}
  {\bibfnamefont {T.}~\bibnamefont {Raubenheimer}},\ and\ \bibinfo {author}
  {\bibfnamefont {D.}~\bibnamefont {Xiang}},\ }\bibfield  {title} {\bibinfo
  {title} {First characterization of coherent optical vortices from harmonic
  undulator radiation},\ }\href@noop {} {\bibfield  {journal} {\bibinfo
  {journal} {Phys. Rev. Lett.}\ }\textbf {\bibinfo {volume} {113}},\ \bibinfo
  {pages} {134803} (\bibinfo {year} {2014})}\BibitemShut {NoStop}%
\bibitem [{\citenamefont {Ribi{\v{c}}}\ \emph {et~al.}(2017)\citenamefont
  {Ribi{\v{c}}}, \citenamefont {R{\"o}sner}, \citenamefont {Gauthier},
  \citenamefont {Allaria}, \citenamefont {D{\"o}ring}, \citenamefont {Foglia},
  \citenamefont {Giannessi}, \citenamefont {Mahne}, \citenamefont {Manfredda},
  \citenamefont {Masciovecchio} \emph {et~al.}}]{PRibivc2017}%
  \BibitemOpen
  \bibfield  {author} {\bibinfo {author} {\bibfnamefont {P.~R.}\ \bibnamefont
  {Ribi{\v{c}}}}, \bibinfo {author} {\bibfnamefont {B.}~\bibnamefont
  {R{\"o}sner}}, \bibinfo {author} {\bibfnamefont {D.}~\bibnamefont
  {Gauthier}}, \bibinfo {author} {\bibfnamefont {E.}~\bibnamefont {Allaria}},
  \bibinfo {author} {\bibfnamefont {F.}~\bibnamefont {D{\"o}ring}}, \bibinfo
  {author} {\bibfnamefont {L.}~\bibnamefont {Foglia}}, \bibinfo {author}
  {\bibfnamefont {L.}~\bibnamefont {Giannessi}}, \bibinfo {author}
  {\bibfnamefont {N.}~\bibnamefont {Mahne}}, \bibinfo {author} {\bibfnamefont
  {M.}~\bibnamefont {Manfredda}}, \bibinfo {author} {\bibfnamefont
  {C.}~\bibnamefont {Masciovecchio}}, \emph {et~al.},\ }\bibfield  {title}
  {\bibinfo {title} {Extreme-ultraviolet vortices from a free-electron laser},\
  }\href@noop {} {\bibfield  {journal} {\bibinfo  {journal} {Phys. Rev. X}\
  }\textbf {\bibinfo {volume} {7}},\ \bibinfo {pages} {031036} (\bibinfo {year}
  {2017})}\BibitemShut {NoStop}%
\bibitem [{\citenamefont {Yu}\ \emph {et~al.}(2000)\citenamefont {Yu},
  \citenamefont {Babzien}, \citenamefont {Ben-Zvi}, \citenamefont {DiMauro},
  \citenamefont {Doyuran}, \citenamefont {Graves}, \citenamefont {Johnson},
  \citenamefont {Krinsky}, \citenamefont {Malone}, \citenamefont {Pogorelsky}
  \emph {et~al.}}]{LYu2000}%
  \BibitemOpen
  \bibfield  {author} {\bibinfo {author} {\bibfnamefont {L.-H.}\ \bibnamefont
  {Yu}}, \bibinfo {author} {\bibfnamefont {M.}~\bibnamefont {Babzien}},
  \bibinfo {author} {\bibfnamefont {I.}~\bibnamefont {Ben-Zvi}}, \bibinfo
  {author} {\bibfnamefont {L.}~\bibnamefont {DiMauro}}, \bibinfo {author}
  {\bibfnamefont {A.}~\bibnamefont {Doyuran}}, \bibinfo {author} {\bibfnamefont
  {W.}~\bibnamefont {Graves}}, \bibinfo {author} {\bibfnamefont
  {E.}~\bibnamefont {Johnson}}, \bibinfo {author} {\bibfnamefont
  {S.}~\bibnamefont {Krinsky}}, \bibinfo {author} {\bibfnamefont
  {R.}~\bibnamefont {Malone}}, \bibinfo {author} {\bibfnamefont
  {I.}~\bibnamefont {Pogorelsky}}, \emph {et~al.},\ }\bibfield  {title}
  {\bibinfo {title} {High-gain harmonic-generation free-electron laser},\
  }\href@noop {} {\bibfield  {journal} {\bibinfo  {journal} {Science}\ }\textbf
  {\bibinfo {volume} {289}},\ \bibinfo {pages} {932} (\bibinfo {year}
  {2000})}\BibitemShut {NoStop}%
\bibitem [{\citenamefont {Billardon}\ \emph {et~al.}(1983)\citenamefont
  {Billardon}, \citenamefont {Elleaume}, \citenamefont {Ortega}, \citenamefont
  {Bazin}, \citenamefont {Bergher}, \citenamefont {Velghe}, \citenamefont
  {Petroff}, \citenamefont {Deacon}, \citenamefont {Robinson},\ and\
  \citenamefont {Madey}}]{MBillardon1983}%
  \BibitemOpen
  \bibfield  {author} {\bibinfo {author} {\bibfnamefont {M.}~\bibnamefont
  {Billardon}}, \bibinfo {author} {\bibfnamefont {P.}~\bibnamefont {Elleaume}},
  \bibinfo {author} {\bibfnamefont {J.~M.}\ \bibnamefont {Ortega}}, \bibinfo
  {author} {\bibfnamefont {C.}~\bibnamefont {Bazin}}, \bibinfo {author}
  {\bibfnamefont {M.}~\bibnamefont {Bergher}}, \bibinfo {author} {\bibfnamefont
  {M.}~\bibnamefont {Velghe}}, \bibinfo {author} {\bibfnamefont
  {Y.}~\bibnamefont {Petroff}}, \bibinfo {author} {\bibfnamefont {D.~A.~G.}\
  \bibnamefont {Deacon}}, \bibinfo {author} {\bibfnamefont {K.~E.}\
  \bibnamefont {Robinson}},\ and\ \bibinfo {author} {\bibfnamefont {J.~M.~J.}\
  \bibnamefont {Madey}},\ }\bibfield  {title} {\bibinfo {title} {First
  operation of a storage-ring free-electron laser},\ }\href@noop {} {\bibfield
  {journal} {\bibinfo  {journal} {Phys. Rev. Lett.}\ }\textbf {\bibinfo
  {volume} {51}},\ \bibinfo {pages} {1652} (\bibinfo {year}
  {1983})}\BibitemShut {NoStop}%
\bibitem [{\citenamefont {Vinokurov}\ \emph {et~al.}(1989)\citenamefont
  {Vinokurov}, \citenamefont {Drobyazko}, \citenamefont {Kulipanov},
  \citenamefont {Litvinenko},\ and\ \citenamefont {Pinayev}}]{NVinokurov1989}%
  \BibitemOpen
  \bibfield  {author} {\bibinfo {author} {\bibfnamefont {N.~A.}\ \bibnamefont
  {Vinokurov}}, \bibinfo {author} {\bibfnamefont {I.~B.}\ \bibnamefont
  {Drobyazko}}, \bibinfo {author} {\bibfnamefont {G.~N.}\ \bibnamefont
  {Kulipanov}}, \bibinfo {author} {\bibfnamefont {V.~N.}\ \bibnamefont
  {Litvinenko}},\ and\ \bibinfo {author} {\bibfnamefont {I.~V.}\ \bibnamefont
  {Pinayev}},\ }\bibfield  {title} {\bibinfo {title} {Lasing in visible and
  ultraviolet regions in an optical klystron installed on the {VEPP-3} storage
  ring},\ }\href@noop {} {\bibfield  {journal} {\bibinfo  {journal} {Rev. Sci.
  Instrum.}\ }\textbf {\bibinfo {volume} {60}},\ \bibinfo {pages} {1435}
  (\bibinfo {year} {1989})}\BibitemShut {NoStop}%
\bibitem [{\citenamefont {Glotin}\ \emph {et~al.}(1993)\citenamefont {Glotin},
  \citenamefont {Chaput}, \citenamefont {Jaroszynski}, \citenamefont
  {Prazeres},\ and\ \citenamefont {Ortega}}]{FGlotin1993}%
  \BibitemOpen
  \bibfield  {author} {\bibinfo {author} {\bibfnamefont {F.}~\bibnamefont
  {Glotin}}, \bibinfo {author} {\bibfnamefont {R.}~\bibnamefont {Chaput}},
  \bibinfo {author} {\bibfnamefont {D.}~\bibnamefont {Jaroszynski}}, \bibinfo
  {author} {\bibfnamefont {R.}~\bibnamefont {Prazeres}},\ and\ \bibinfo
  {author} {\bibfnamefont {J.-M.}\ \bibnamefont {Ortega}},\ }\bibfield  {title}
  {\bibinfo {title} {Infrared subpicosecond laser pulses with a free-electron
  laser},\ }\href@noop {} {\bibfield  {journal} {\bibinfo  {journal} {Phys.
  Rev. Lett.}\ }\textbf {\bibinfo {volume} {71}},\ \bibinfo {pages} {2587}
  (\bibinfo {year} {1993})}\BibitemShut {NoStop}%
\bibitem [{\citenamefont {Takano}\ \emph {et~al.}(1993)\citenamefont {Takano},
  \citenamefont {Hama},\ and\ \citenamefont {Isoyama}}]{STakano1993}%
  \BibitemOpen
  \bibfield  {author} {\bibinfo {author} {\bibfnamefont {S.}~\bibnamefont
  {Takano}}, \bibinfo {author} {\bibfnamefont {H.}~\bibnamefont {Hama}},\ and\
  \bibinfo {author} {\bibfnamefont {G.}~\bibnamefont {Isoyama}},\ }\bibfield
  {title} {\bibinfo {title} {Lasing of a free electron laser in the visible on
  the {UVSOR} storage ring},\ }\href@noop {} {\bibfield  {journal} {\bibinfo
  {journal} {Nucl. Instrum. Methods Phys. Res., Sect. A}\ }\textbf {\bibinfo
  {volume} {331}},\ \bibinfo {pages} {20} (\bibinfo {year} {1993})}\BibitemShut
  {NoStop}%
\bibitem [{\citenamefont {Neil}\ \emph {et~al.}(2000)\citenamefont {Neil},
  \citenamefont {Bohn}, \citenamefont {Benson}, \citenamefont {Biallas},
  \citenamefont {Douglas}, \citenamefont {Dylla}, \citenamefont {Evans},
  \citenamefont {Fugitt}, \citenamefont {Grippo}, \citenamefont {Gubeli} \emph
  {et~al.}}]{GNeil2000}%
  \BibitemOpen
  \bibfield  {author} {\bibinfo {author} {\bibfnamefont {G.~R.}\ \bibnamefont
  {Neil}}, \bibinfo {author} {\bibfnamefont {C.}~\bibnamefont {Bohn}}, \bibinfo
  {author} {\bibfnamefont {S.}~\bibnamefont {Benson}}, \bibinfo {author}
  {\bibfnamefont {G.}~\bibnamefont {Biallas}}, \bibinfo {author} {\bibfnamefont
  {D.}~\bibnamefont {Douglas}}, \bibinfo {author} {\bibfnamefont
  {H.}~\bibnamefont {Dylla}}, \bibinfo {author} {\bibfnamefont
  {R.}~\bibnamefont {Evans}}, \bibinfo {author} {\bibfnamefont
  {J.}~\bibnamefont {Fugitt}}, \bibinfo {author} {\bibfnamefont
  {A.}~\bibnamefont {Grippo}}, \bibinfo {author} {\bibfnamefont
  {J.}~\bibnamefont {Gubeli}}, \emph {et~al.},\ }\bibfield  {title} {\bibinfo
  {title} {Sustained kilowatt lasing in a free-electron laser with same-cell
  energy recovery},\ }\href@noop {} {\bibfield  {journal} {\bibinfo  {journal}
  {Phys. Rev. Lett.}\ }\textbf {\bibinfo {volume} {84}},\ \bibinfo {pages}
  {662} (\bibinfo {year} {2000})}\BibitemShut {NoStop}%
\bibitem [{\citenamefont {Litvinenko}\ \emph {et~al.}(2001)\citenamefont
  {Litvinenko}, \citenamefont {Park}, \citenamefont {Pinayev},\ and\
  \citenamefont {Wu}}]{VLitvinenko2001}%
  \BibitemOpen
  \bibfield  {author} {\bibinfo {author} {\bibfnamefont {V.~N.}\ \bibnamefont
  {Litvinenko}}, \bibinfo {author} {\bibfnamefont {S.~H.}\ \bibnamefont
  {Park}}, \bibinfo {author} {\bibfnamefont {I.~V.}\ \bibnamefont {Pinayev}},\
  and\ \bibinfo {author} {\bibfnamefont {Y.}~\bibnamefont {Wu}},\ }\bibfield
  {title} {\bibinfo {title} {Operation of the {OK-4/Duke} storage ring {FEL}
  below 200 nm},\ }\href@noop {} {\bibfield  {journal} {\bibinfo  {journal}
  {Nucl. Instrum. Methods Phys. Res., Sect. A}\ }\textbf {\bibinfo {volume}
  {475}},\ \bibinfo {pages} {195} (\bibinfo {year} {2001})}\BibitemShut
  {NoStop}%
\bibitem [{\citenamefont {Shevchenko}\ \emph {et~al.}(2016)\citenamefont
  {Shevchenko}, \citenamefont {Arbuzov}, \citenamefont {Vinokurov},
  \citenamefont {Vobly}, \citenamefont {Volkov}, \citenamefont {Getmanov},
  \citenamefont {Gorbachev}, \citenamefont {Davidyuk}, \citenamefont
  {Deychuly}, \citenamefont {Dementyev} \emph {et~al.}}]{OShevchenko2016}%
  \BibitemOpen
  \bibfield  {author} {\bibinfo {author} {\bibfnamefont {O.~A.}\ \bibnamefont
  {Shevchenko}}, \bibinfo {author} {\bibfnamefont {V.~S.}\ \bibnamefont
  {Arbuzov}}, \bibinfo {author} {\bibfnamefont {N.~A.}\ \bibnamefont
  {Vinokurov}}, \bibinfo {author} {\bibfnamefont {P.~D.}\ \bibnamefont
  {Vobly}}, \bibinfo {author} {\bibfnamefont {V.~N.}\ \bibnamefont {Volkov}},
  \bibinfo {author} {\bibfnamefont {Y.~V.}\ \bibnamefont {Getmanov}}, \bibinfo
  {author} {\bibfnamefont {Y.~I.}\ \bibnamefont {Gorbachev}}, \bibinfo {author}
  {\bibfnamefont {I.~V.}\ \bibnamefont {Davidyuk}}, \bibinfo {author}
  {\bibfnamefont {O.~I.}\ \bibnamefont {Deychuly}}, \bibinfo {author}
  {\bibfnamefont {E.~N.}\ \bibnamefont {Dementyev}}, \emph {et~al.},\
  }\bibfield  {title} {\bibinfo {title} {The {Novosibirsk Free Electron
  Laser}--unique source of terahertz and infrared coherent radiation},\
  }\href@noop {} {\bibfield  {journal} {\bibinfo  {journal} {Phys. Procedia}\
  }\textbf {\bibinfo {volume} {84}},\ \bibinfo {pages} {13} (\bibinfo {year}
  {2016})}\BibitemShut {NoStop}%
\bibitem [{\citenamefont {Wu}\ \emph {et~al.}(2006)\citenamefont {Wu},
  \citenamefont {Vinokurov}, \citenamefont {Mikhailov}, \citenamefont {Li},\
  and\ \citenamefont {Popov}}]{wu2006high}%
  \BibitemOpen
  \bibfield  {author} {\bibinfo {author} {\bibfnamefont {Y.~K.}\ \bibnamefont
  {Wu}}, \bibinfo {author} {\bibfnamefont {N.~A.}\ \bibnamefont {Vinokurov}},
  \bibinfo {author} {\bibfnamefont {S.}~\bibnamefont {Mikhailov}}, \bibinfo
  {author} {\bibfnamefont {J.}~\bibnamefont {Li}},\ and\ \bibinfo {author}
  {\bibfnamefont {V.}~\bibnamefont {Popov}},\ }\bibfield  {title} {\bibinfo
  {title} {High-gain lasing and polarization switch with a distributed
  optical-klystron free-electron laser},\ }\href@noop {} {\bibfield  {journal}
  {\bibinfo  {journal} {Phys. Rev. Lett.}\ }\textbf {\bibinfo {volume} {96}},\
  \bibinfo {pages} {224801} (\bibinfo {year} {2006})}\BibitemShut {NoStop}%
\bibitem [{\citenamefont {Vinokurov}\ and\ \citenamefont
  {Skrinsky}()}]{vinokurov1977report}%
  \BibitemOpen
  \bibfield  {author} {\bibinfo {author} {\bibfnamefont {N.~A.}\ \bibnamefont
  {Vinokurov}}\ and\ \bibinfo {author} {\bibfnamefont {A.~N.}\ \bibnamefont
  {Skrinsky}},\ }\bibinfo {note} {{Budker Institute of Nuclear Physics,
  Novosibirsk Report No. INP 77-59, 1977 (to be published)}}\BibitemShut
  {NoStop}%
\bibitem [{\citenamefont {Litvinenko}\ \emph {et~al.}(1998)\citenamefont
  {Litvinenko} \emph {et~al.}}]{litvinenko1998first}%
  \BibitemOpen
  \bibfield  {author} {\bibinfo {author} {\bibfnamefont {V.~N.}\ \bibnamefont
  {Litvinenko}} \emph {et~al.},\ }\bibfield  {title} {\bibinfo {title} {{First
  UV/visible lasing with the OK-4/Duke storage ring FEL}},\ }\href@noop {}
  {\bibfield  {journal} {\bibinfo  {journal} {Nucl. Instrum. Methods Phys.
  Res., Sect. A}\ }\textbf {\bibinfo {volume} {407}},\ \bibinfo {pages} {8}
  (\bibinfo {year} {1998})}\BibitemShut {NoStop}%
\bibitem [{\citenamefont {Pereira}\ \emph {et~al.}(1998)\citenamefont
  {Pereira}, \citenamefont {Willemsen}, \citenamefont {van Exter},\ and\
  \citenamefont {Woerdman}}]{pereira1998pinning}%
  \BibitemOpen
  \bibfield  {author} {\bibinfo {author} {\bibfnamefont {S.~F.}\ \bibnamefont
  {Pereira}}, \bibinfo {author} {\bibfnamefont {M.~B.}\ \bibnamefont
  {Willemsen}}, \bibinfo {author} {\bibfnamefont {M.~P.}\ \bibnamefont {van
  Exter}},\ and\ \bibinfo {author} {\bibfnamefont {J.~P.}\ \bibnamefont
  {Woerdman}},\ }\bibfield  {title} {\bibinfo {title} {Pinning of daisy modes
  in optically pumped vertical-cavity surface-emitting lasers},\ }\href@noop {}
  {\bibfield  {journal} {\bibinfo  {journal} {Appl. Phys. Lett.}\ }\textbf
  {\bibinfo {volume} {73}},\ \bibinfo {pages} {2239} (\bibinfo {year}
  {1998})}\BibitemShut {NoStop}%
\bibitem [{\citenamefont {Chen}\ \emph {et~al.}(2001)\citenamefont {Chen},
  \citenamefont {Lan},\ and\ \citenamefont {Wang}}]{chen2001generation}%
  \BibitemOpen
  \bibfield  {author} {\bibinfo {author} {\bibfnamefont {Y.~F.}\ \bibnamefont
  {Chen}}, \bibinfo {author} {\bibfnamefont {Y.~P.}\ \bibnamefont {Lan}},\ and\
  \bibinfo {author} {\bibfnamefont {S.~C.}\ \bibnamefont {Wang}},\ }\bibfield
  {title} {\bibinfo {title} {{Generation of Laguerre--Gaussian modes in
  fiber-coupled laser diode end-pumped lasers}},\ }\href@noop {} {\bibfield
  {journal} {\bibinfo  {journal} {Appl. Phys. B}\ }\textbf {\bibinfo {volume}
  {72}},\ \bibinfo {pages} {167} (\bibinfo {year} {2001})}\BibitemShut
  {NoStop}%
\bibitem [{\citenamefont {Naidoo}\ \emph {et~al.}(2012)\citenamefont {Naidoo},
  \citenamefont {A{\"\i}t-Ameur}, \citenamefont {Brunel},\ and\ \citenamefont
  {Forbes}}]{naidoo2012intra}%
  \BibitemOpen
  \bibfield  {author} {\bibinfo {author} {\bibfnamefont {D.}~\bibnamefont
  {Naidoo}}, \bibinfo {author} {\bibfnamefont {K.}~\bibnamefont
  {A{\"\i}t-Ameur}}, \bibinfo {author} {\bibfnamefont {M.}~\bibnamefont
  {Brunel}},\ and\ \bibinfo {author} {\bibfnamefont {A.}~\bibnamefont
  {Forbes}},\ }\bibfield  {title} {\bibinfo {title} {{Intra-cavity generation
  of superpositions of Laguerre--Gaussian beams}},\ }\href@noop {} {\bibfield
  {journal} {\bibinfo  {journal} {Appl. Phys. B}\ }\textbf {\bibinfo {volume}
  {106}},\ \bibinfo {pages} {683} (\bibinfo {year} {2012})}\BibitemShut
  {NoStop}%
\bibitem [{\citenamefont {Lin}\ \emph {et~al.}(2014)\citenamefont {Lin},
  \citenamefont {Daniel},\ and\ \citenamefont {Clarkson}}]{lin2014controlling}%
  \BibitemOpen
  \bibfield  {author} {\bibinfo {author} {\bibfnamefont {D.}~\bibnamefont
  {Lin}}, \bibinfo {author} {\bibfnamefont {J.~M.~O.}\ \bibnamefont {Daniel}},\
  and\ \bibinfo {author} {\bibfnamefont {W.~A.}\ \bibnamefont {Clarkson}},\
  }\bibfield  {title} {\bibinfo {title} {{Controlling the handedness of
  directly excited Laguerre--Gaussian modes in a solid-state laser}},\
  }\href@noop {} {\bibfield  {journal} {\bibinfo  {journal} {Opt. Lett.}\
  }\textbf {\bibinfo {volume} {39}},\ \bibinfo {pages} {3903} (\bibinfo {year}
  {2014})}\BibitemShut {NoStop}%
\bibitem [{\citenamefont {Siegman}(1990)}]{siegman1990new}%
  \BibitemOpen
  \bibfield  {author} {\bibinfo {author} {\bibfnamefont {A.~E.}\ \bibnamefont
  {Siegman}},\ }\bibfield  {title} {\bibinfo {title} {New developments in laser
  resonators},\ }\href@noop {} {\bibfield  {journal} {\bibinfo  {journal}
  {Proc. SPIE}\ }\textbf {\bibinfo {volume} {1224}},\ \bibinfo {pages} {2}
  (\bibinfo {year} {1990})}\BibitemShut {NoStop}%
\bibitem [{\citenamefont {Liu}\ \emph {et~al.}(2020)\citenamefont {Liu},
  \citenamefont {Yan}, \citenamefont {Hao},\ and\ \citenamefont
  {Wu}}]{liu2020phaseret}%
  \BibitemOpen
  \bibfield  {author} {\bibinfo {author} {\bibfnamefont {P.}~\bibnamefont
  {Liu}}, \bibinfo {author} {\bibfnamefont {J.}~\bibnamefont {Yan}}, \bibinfo
  {author} {\bibfnamefont {H.}~\bibnamefont {Hao}},\ and\ \bibinfo {author}
  {\bibfnamefont {Y.~K.}\ \bibnamefont {Wu}},\ }\bibfield  {title} {\bibinfo
  {title} {Phase retrieval for short wavelength orbital angular momentum beams
  using knife-edge diffraction},\ }\href@noop {} {\bibfield  {journal}
  {\bibinfo  {journal} {Opt. Commun.}\ }\textbf {\bibinfo {volume} {474}},\
  \bibinfo {pages} {126077} (\bibinfo {year} {2020})}\BibitemShut {NoStop}%
\bibitem [{\citenamefont {Billardon}(1990)}]{billardon1990storage}%
  \BibitemOpen
  \bibfield  {author} {\bibinfo {author} {\bibfnamefont {M.}~\bibnamefont
  {Billardon}},\ }\bibfield  {title} {\bibinfo {title} {Storage ring
  free-electron laser and chaos},\ }\href@noop {} {\bibfield  {journal}
  {\bibinfo  {journal} {Phys. Rev. Lett.}\ }\textbf {\bibinfo {volume} {65}},\
  \bibinfo {pages} {713} (\bibinfo {year} {1990})}\BibitemShut {NoStop}%
\bibitem [{\citenamefont {Billardon}\ \emph {et~al.}(1992)\citenamefont
  {Billardon}, \citenamefont {Garzella},\ and\ \citenamefont
  {Couprie}}]{billardon1992saturation}%
  \BibitemOpen
  \bibfield  {author} {\bibinfo {author} {\bibfnamefont {M.}~\bibnamefont
  {Billardon}}, \bibinfo {author} {\bibfnamefont {D.}~\bibnamefont
  {Garzella}},\ and\ \bibinfo {author} {\bibfnamefont {M.~E.}\ \bibnamefont
  {Couprie}},\ }\bibfield  {title} {\bibinfo {title} {Saturation mechanism for
  a storage-ring free-electron laser},\ }\href@noop {} {\bibfield  {journal}
  {\bibinfo  {journal} {Phys. Rev. Lett.}\ }\textbf {\bibinfo {volume} {69}},\
  \bibinfo {pages} {2368} (\bibinfo {year} {1992})}\BibitemShut {NoStop}%
\bibitem [{\citenamefont {Wu}\ \emph {et~al.}(2015)\citenamefont {Wu},
  \citenamefont {Yan}, \citenamefont {Hao}, \citenamefont {Li}, \citenamefont
  {Mikhailov}, \citenamefont {Popov}, \citenamefont {Vinokurov}, \citenamefont
  {Huang},\ and\ \citenamefont {Wu}}]{YWu2015}%
  \BibitemOpen
  \bibfield  {author} {\bibinfo {author} {\bibfnamefont {Y.~K.}\ \bibnamefont
  {Wu}}, \bibinfo {author} {\bibfnamefont {J.}~\bibnamefont {Yan}}, \bibinfo
  {author} {\bibfnamefont {H.}~\bibnamefont {Hao}}, \bibinfo {author}
  {\bibfnamefont {J.~Y.}\ \bibnamefont {Li}}, \bibinfo {author} {\bibfnamefont
  {S.~F.}\ \bibnamefont {Mikhailov}}, \bibinfo {author} {\bibfnamefont {V.~G.}\
  \bibnamefont {Popov}}, \bibinfo {author} {\bibfnamefont {N.~A.}\ \bibnamefont
  {Vinokurov}}, \bibinfo {author} {\bibfnamefont {S.}~\bibnamefont {Huang}},\
  and\ \bibinfo {author} {\bibfnamefont {J.}~\bibnamefont {Wu}},\ }\bibfield
  {title} {\bibinfo {title} {Widely tunable two-color free-electron laser on a
  storage ring},\ }\href@noop {} {\bibfield  {journal} {\bibinfo  {journal}
  {Phys. Rev. Lett.}\ }\textbf {\bibinfo {volume} {115}},\ \bibinfo {pages}
  {184801} (\bibinfo {year} {2015})}\BibitemShut {NoStop}%
\bibitem [{\citenamefont {Yan}\ \emph {et~al.}(2019)\citenamefont {Yan} \emph
  {et~al.}}]{yan2019precision}%
  \BibitemOpen
  \bibfield  {author} {\bibinfo {author} {\bibfnamefont {J.}~\bibnamefont
  {Yan}} \emph {et~al.},\ }\bibfield  {title} {\bibinfo {title} {Precision
  control of gamma-ray polarization using a crossed helical undulator
  free-electron laser},\ }\href@noop {} {\bibfield  {journal} {\bibinfo
  {journal} {Nat. Photon.}\ }\textbf {\bibinfo {volume} {13}},\ \bibinfo
  {pages} {629} (\bibinfo {year} {2019})}\BibitemShut {NoStop}%
\bibitem [{\citenamefont {Kim}\ \emph {et~al.}(2008)\citenamefont {Kim},
  \citenamefont {Shvyd’ko},\ and\ \citenamefont {Reiche}}]{kim2008proposal}%
  \BibitemOpen
  \bibfield  {author} {\bibinfo {author} {\bibfnamefont {K.-J.}\ \bibnamefont
  {Kim}}, \bibinfo {author} {\bibfnamefont {Y.}~\bibnamefont {Shvyd’ko}},\
  and\ \bibinfo {author} {\bibfnamefont {S.}~\bibnamefont {Reiche}},\
  }\bibfield  {title} {\bibinfo {title} {A proposal for an x-ray free-electron
  laser oscillator with an energy-recovery linac},\ }\href@noop {} {\bibfield
  {journal} {\bibinfo  {journal} {Phys. Rev. Lett.}\ }\textbf {\bibinfo
  {volume} {100}},\ \bibinfo {pages} {244802} (\bibinfo {year}
  {2008})}\BibitemShut {NoStop}%
\bibitem [{\citenamefont {Stupakov}\ and\ \citenamefont
  {Zolotorev}(2014)}]{GStupakov2014}%
  \BibitemOpen
  \bibfield  {author} {\bibinfo {author} {\bibfnamefont {G.}~\bibnamefont
  {Stupakov}}\ and\ \bibinfo {author} {\bibfnamefont {M.~S.}\ \bibnamefont
  {Zolotorev}},\ }\href@noop {} {\emph {\bibinfo {title} {FEL oscillator for
  {EUV} lithography}}},\ \bibinfo {type} {Tech. Rep.}\ (\bibinfo  {institution}
  {SLAC National Accelerator Lab., Menlo Park, CA (United States)},\ \bibinfo
  {year} {2014})\BibitemShut {NoStop}%
\bibitem [{\citenamefont {Benson}\ \emph {et~al.}(2010)\citenamefont {Benson},
  \citenamefont {Douglas}, \citenamefont {Evtushenko}, \citenamefont {Gubeli},
  \citenamefont {Hannon}, \citenamefont {Jordan}, \citenamefont {Klopf},
  \citenamefont {Neil}, \citenamefont {Shinn}, \citenamefont {Tennant} \emph
  {et~al.}}]{SBenson2010}%
  \BibitemOpen
  \bibfield  {author} {\bibinfo {author} {\bibfnamefont {S.~V.}\ \bibnamefont
  {Benson}}, \bibinfo {author} {\bibfnamefont {D.}~\bibnamefont {Douglas}},
  \bibinfo {author} {\bibfnamefont {P.}~\bibnamefont {Evtushenko}}, \bibinfo
  {author} {\bibfnamefont {J.}~\bibnamefont {Gubeli}}, \bibinfo {author}
  {\bibfnamefont {F.~E.}\ \bibnamefont {Hannon}}, \bibinfo {author}
  {\bibfnamefont {K.}~\bibnamefont {Jordan}}, \bibinfo {author} {\bibfnamefont
  {J.~M.}\ \bibnamefont {Klopf}}, \bibinfo {author} {\bibfnamefont {G.~R.}\
  \bibnamefont {Neil}}, \bibinfo {author} {\bibfnamefont {M.~D.}\ \bibnamefont
  {Shinn}}, \bibinfo {author} {\bibfnamefont {C.}~\bibnamefont {Tennant}},
  \emph {et~al.},\ }\bibfield  {title} {\bibinfo {title} {The {JLAMP VUV}/soft
  x-ray user facility at {Jefferson Laboratory}},\ }in\ \href@noop {} {\emph
  {\bibinfo {booktitle} {{Proceedings, 1st International Particle Accelerator
  Conference (IPAC 2010): Kyoto, Japan, 2010}}}}\ (\bibinfo {year} {2010})\ p.\
  \bibinfo {pages} {2302}\BibitemShut {NoStop}%
\bibitem [{\citenamefont {Socol}\ \emph {et~al.}(2011)\citenamefont {Socol},
  \citenamefont {Kulipanov}, \citenamefont {Matveenko}, \citenamefont
  {Shevchenko},\ and\ \citenamefont {Vinokurov}}]{YSocol2011}%
  \BibitemOpen
  \bibfield  {author} {\bibinfo {author} {\bibfnamefont {Y.}~\bibnamefont
  {Socol}}, \bibinfo {author} {\bibfnamefont {G.}~\bibnamefont {Kulipanov}},
  \bibinfo {author} {\bibfnamefont {A.}~\bibnamefont {Matveenko}}, \bibinfo
  {author} {\bibfnamefont {O.}~\bibnamefont {Shevchenko}},\ and\ \bibinfo
  {author} {\bibfnamefont {N.}~\bibnamefont {Vinokurov}},\ }\bibfield  {title}
  {\bibinfo {title} {Compact 13.5-nm free-electron laser for extreme
  ultraviolet lithography},\ }\href@noop {} {\bibfield  {journal} {\bibinfo
  {journal} {Phys. Rev. ST Accel. Beams}\ }\textbf {\bibinfo {volume} {14}},\
  \bibinfo {pages} {040702} (\bibinfo {year} {2011})}\BibitemShut {NoStop}%
\bibitem [{\citenamefont {Jentschura}\ and\ \citenamefont
  {Serbo}(2011)}]{jentschura2011generation}%
  \BibitemOpen
  \bibfield  {author} {\bibinfo {author} {\bibfnamefont {U.~D.}\ \bibnamefont
  {Jentschura}}\ and\ \bibinfo {author} {\bibfnamefont {V.~G.}\ \bibnamefont
  {Serbo}},\ }\bibfield  {title} {\bibinfo {title} {{Generation of high-energy
  photons with large orbital angular momentum by Compton backscattering}},\
  }\href@noop {} {\bibfield  {journal} {\bibinfo  {journal} {Phys. Rev. Lett.}\
  }\textbf {\bibinfo {volume} {106}},\ \bibinfo {pages} {013001} (\bibinfo
  {year} {2011})}\BibitemShut {NoStop}%
\bibitem [{\citenamefont {Petrillo}\ \emph {et~al.}(2016)\citenamefont
  {Petrillo}, \citenamefont {Dattoli}, \citenamefont {Drebot},\ and\
  \citenamefont {Nguyen}}]{petrillo2016compton}%
  \BibitemOpen
  \bibfield  {author} {\bibinfo {author} {\bibfnamefont {V.}~\bibnamefont
  {Petrillo}}, \bibinfo {author} {\bibfnamefont {G.}~\bibnamefont {Dattoli}},
  \bibinfo {author} {\bibfnamefont {I.}~\bibnamefont {Drebot}},\ and\ \bibinfo
  {author} {\bibfnamefont {F.}~\bibnamefont {Nguyen}},\ }\bibfield  {title}
  {\bibinfo {title} {Compton scattered x-gamma rays with orbital momentum},\
  }\href@noop {} {\bibfield  {journal} {\bibinfo  {journal} {Phys. Rev. Lett.}\
  }\textbf {\bibinfo {volume} {117}},\ \bibinfo {pages} {123903} (\bibinfo
  {year} {2016})}\BibitemShut {NoStop}%
\end{thebibliography}
\end{document}